\documentclass[prd,showkeys,twocolumn,preprintnumbers,amsmath,amssymb,superscriptaddress,floatfix,aps,nofootinbib]{revtex4}
\usepackage{graphicx,color,dcolumn,booktabs,bm}
\usepackage{amsmath}
\usepackage{graphicx}
\usepackage{longtable,lscape}
\usepackage{txfonts}
\usepackage{textcomp}
\usepackage{overpic}
\usepackage{epsfig}
\usepackage{amssymb}
\usepackage{rotating}
\usepackage{epstopdf}
\usepackage{appendix}
\usepackage{indentfirst}
\usepackage{feynmf}   
\usepackage{slashed}  
\usepackage{cases}
\usepackage{color}
\usepackage{multirow}
\usepackage{graphicx,color,dcolumn,booktabs,bm}
\usepackage{cases}
\usepackage{array}
\usepackage{changes}
\usepackage{float}
\usepackage{placeins}

\graphicspath{{Figures/}} %

\usepackage[colorlinks, citecolor=blue,anchorcolor=red,menucolor=red, linkcolor=red,filecolor=red,runcolor=red,urlcolor=blue,frenchlinks=red]{hyperref}
\usepackage{epstopdf}
\newcommand{\MeV}{\ensuremath{\mathrm{MeV}}}

\begin{document}
\title{An analysis on doubly bottom molecular tetraquarks composed of $H_{(s)}$ and $T_{(s)}$ doublets }

\author{Jun-Chao Su}
\affiliation{School of Physics and Electronics, Henan University,  Kaifeng 475004, China}

\author{Qing-Fu Song}
\affiliation{School of Physics, Central South University, Changsha 410083, China}

\author{Qi-Fang L\"{u}}\email{lvqifang@hunnu.edu.cn}

\affiliation{Department of Physics, Hunan Normal University, Changsha 410081, China}
\affiliation{Key Laboratory of Low-Dimensional Quantum Structures and Quantum Control of Ministry of Education, Changsha 410081, China}
\affiliation{Key Laboratory for Matter Microstructure and Function of Hunan Province, Hunan Normal University, Changsha 410081, China}

\author{Jingya Zhu}
\email{zhujy@henu.edu.cn}
\affiliation{School of Physics and Electronics, Henan University,  Kaifeng 475004, China}

\begin{abstract}
In this work, we investigate the doubly bottom $H_{(s)}\bar{T}_{(s)}$  and $H_{(s)}T_{(s)}$ systems by adopting the one-boson-exchange model, where $H_{(s)}$ and $T_{(s)}$ represent $S$-wave $B^{(*)}_{(s)}$ and $P$-wave $B^{(*)}_{(s)1,2}$ doublets, respectively. For the $H\bar{T}$ systems, we predict some loosely bound states in the $I(J^{PC})=0(1^{-\pm})$ $B\bar{B}_{1}$, $I(J^{PC})=0(2^{-\pm})$ $B\bar{B}_{2}^{*}$, $I(J^{PC})=0(1^{-\pm})$ $B^*\bar{B}_{1}$ and $I(J^{PC})=0(2^{-\pm})$ $B^*\bar{B}_{2}^{*}$ channels, which are the most promising hidden bottom molecular tetraquarks. For the $HT$ systems, the $B^*B_1$ channels with quantum numbers $I(J^P) = 0(1^{-}), 0(2^{-})$ and the $B^*B_2^*$ channels with $I(J^P) = 0(2^{-})$ are also likely candidates for forming molecular tetraquarks. In contrast, no molecular candidates have been identified in the bottom-strange sectors. One can hope that our predictions will provide valuable insights to the LHCb and Belle II Collaborations as they continue to explore this fascinating field through experimental research.
\end{abstract}


\maketitle
\section{Introduction}

In addition to conventional mesons and baryons, quantum chromodynamics (QCD) predicts the existence of exotic hadronic states, including tetraquarks, pentaquarks, hexaquarks, hybrid states, and glueballs. Subsequent to the discovery of the $X(3872)$ particle by the Belle Collaboration in 2003~\cite{Belle:2003nnu}, a plethora of exotic hadronic states have been unearthed through experiments conducted at prominent facilities, including charmonium-like and bottomonium-like $XYZ$ states, tetraquark states, and pentaquark states $P_{c(s)}$, which has prompted extensive theoretical investigations. The proximity of numerous new hadronic states to hadron-hadron thresholds suggests their interpretation as molecular states~\cite{Liu:2019zoy, Wang:2025sic,Chen:2016qju,Liu:2024uxn,Hosaka:2016pey,Lebed:2016hpi,Ali:2017jda,
Esposito:2016noz,Chen:2022asf,Dong:2017gaw,Guo:2017jvc,Olsen:2017bmm,
Karliner:2017qhf,Brambilla:2019esw,Bicudo:2022cqi,Mai:2022eur,
Barabanov:2020jvn,Brambilla:2022ura,Ali:2019roi,Esposito:2014rxa,Ali:2018bdc}. The investigations of molecular states can offer significant insights into hadronic spectra and structures, thereby enhancing our comprehension of non-perturbative QCD phenomena. Consequently, the identification of genuine molecular states among existing candidates and the prediction of new ones for future experimental investigations represent a crucial step in hadron physics research.

The charm sector has emerged as a fertile ground for molecular tetraquark candidates. The $X(3872)$, hypothesized as a $D\bar{D}^*+\text{h.c.}$ molecule near the $D\bar{D}^*$ threshold~\cite{Belle:2003nnu}, has stimulated substantial theoretical investigations. In 2013, the BESIII and Belle Collaborations simultaneously discovered $Z_c(3900)$~\cite{BESIII:2013ris, Belle:2013yex}, located marginally above the $D\bar{D}^*$ threshold, establishing it as the isospin partner of $X(3872)$. 
Subsequently, the discovery of $Z_{cs}(4000)$ with the $c\bar{c}u\bar{s}$ configuration in the $J/\psi K^+$ decay channels marked a significant advance. 
Subsequent to this, BESIII identified $Z_c(4020)$~\cite{BESIII:2013ouc} and Belle detected $X(4013)$~\cite{Belle:2021nuv}. Both of these exhibit characteristics predicted by heavy-quark spin symmetry (HQSS). Recent LHCb results from $B^+\to D^{*\pm}D^\mp K^+$ decays revealed the $\chi_{c1}(4010)$ structure with potential exotic contributions~\cite{LHCb:2024vfz}.
The experimental landscape extends beyond the positive-parity states, with $1^{--}$ molecular candidates emerging across multiple facilities. 
The $Y(4260)$, initially observed by BaBar in $e^+e^-\to\pi^+\pi^-J/\psi$~\cite{BaBar:2005hhc}, was subsequently confirmed by independent measurements by Belle~\cite{Belle:2007dxy} and BESIII~\cite{BESIII:2013ris}. 
Subsequent high-precision measurements have refined its mass to $4222~\rm{MeV}$, thus necessitating its reclassification as the $\psi(4230)$ state. 
Concurrent observations by the BaBar and Belle Collaborations substantiated the existence of $Y(4360)$ in the $\pi^+\pi^-\psi(2S)$ channels~\cite{BaBar:2006ait, BaBar:2012hpr, Belle:2014wyt}, while the $e^+e^-\to D_s^+D_{s1}(2536)^-$ analysis conducted by Belle yielded the discovery of the $Y(4626)$ state~\cite{Belle:2019qoi}.
A breakthrough emerged in 2019 when the LHCb Collaboration identified the $T^+_{cc}$ tetraquark with $pp$ collision data~\cite{LHCb:2021vvq}. 
This $ccu\bar{d}$ configuration manifests as a narrow resonance ($m \approx 3875~\rm{MeV}$) in the $D^0D^0\pi^+$ spectrum. 
Comprehensive spectral analysis determined its mass relation to the $D^{*+}D^0$ threshold, providing conclusive evidence for its exotic nature~\cite{LHCb:2021auc}.

Theoretically, the $X(3872)$, $Z_c(3900)$, $Z_{cs}(4000)$, $Z_c(4020)$, $X(4013)$, $\chi_{c1}(4010)$, and $Z_c(4430)$ are {typically} interpreted as hadronic molecules containing exotic configurations of $S$-wave charmed mesons. Besides the molecules composed of two $S$-wave mesons, the $Y(4230)$ has been extensively analyzed as a $D\bar{D}_1(2420)+\text{h.c.}$ molecular configuration through multiple theoretical approaches~\cite{Close:2009ag,Close:2010wq}. 
Furthermore, the $Y(4360)$ and $Y(4390)$ resonances demonstrate intriguing systematics: Both resonances reside below the $D^*D_1(2420)+\text{h.c.}$ threshold, exhibiting characteristics compatible with molecular tetraquark configurations~\cite{Ding:2008gr,Cleven:2013mka,Wang:2013kra}. The $Y(4626)$ state presents particular interest, interpreted as a superposition state of $D_s\bar{D}_{s1}(2536)$ and $D_s^*\bar{D}_{s1}(2536)$ configurations~\cite{He:2019csk}. {In addition to the molecular interpretation, there are many other possible explanations for these newly observed hadronic states in experiments, such as compact tetraquarks and hybrid states~\cite{Chen:2016qju,Liu:2019zoy,Chen:2022asf,Ali:2017jda,Esposito:2016noz,Guo:2017jvc,Maiani:2004vq,Deng:2014gqa,Lu:2016cwr,Kim:2016tys,Chen:2017dpy,Anwar:2018sol,Giron:2019bcs,Berwein:2015vca,Miyamoto:2018zfr,Cleven:2015era}. }Current research on these new hadronic states is still ongoing, and how to distinguish between these different explanations in combination with experimental data remains a hot topic in hadron physics. In this work, we focus exclusively on the study of loosely bound molecular states.

In sharp contrast to the well-characterized charmonium-like spectrum, the bottomonium-like sector remains largely unexplored territory. To date, only a few bottomonium-like states like $Z_b(10610)$, $Z_b(10650)$, and $\Upsilon(10753)$ have been experimentally confirmed by the Belle Collaboration~\cite{Belle:2011aa,Belle:2015upu,Belle-II:2024mjm}. Current theoretical frameworks assign $Z_b(10610)$ and $Z_b(10650)$ to $B\bar{B}$ and $B\bar{B}^*+\text{h.c.}$ molecular configurations, respectively, while $\Upsilon(10753)$ may exhibit characteristics consistent with an exotic state. Despite many theoretical investigations of $S$-wave bottom-strange mesons~\cite{Bondar:2011ev,Sun:2011uh,Mehen:2011yh,Cleven:2011gp,Dias:2014pva,Li:2012wf,Yang:2011rp,Zhang:2011jja,Wang:2013daa,Wang:2014gwa,Dong:2012hc,Ohkoda:2013cea,Li:2012as,Cleven:2013sq,Li:2012uc,Li:2014pfa,Xiao:2017uve,Wu:2018xaa,Wu:2020edh,Voloshin:2011qa,Cheng:2023vyv}, the $H_{(s)}\bar{T}_{(s)}$ and $H_{(s)}T_{(s)}$ systems remain conspicuously underdeveloped. 
Here, the $H_{(s)}$ denotes the $S$-wave heavy-light mesons, and $T_{(s)}$ stands for the $P$-wave doublets with light spin $j=3/2$. Notably, analogous to charmonium systems, the bottom sector should host $1^{--}$ molecular states composed of $H$ and $T$ mesons, which could provide critical insights into the structural properties of $\Upsilon(10753)$, $\Upsilon(10860)$, and $\Upsilon(11020)$. This compelling theoretical landscape necessitates systematic investigations of $H_{(s)}T_{(s)}$ and $T_{(s)}\bar{T}_{(s)}$ configurations employing  methodologies consistent with the studies of charmonium-like states.

The current investigation conducts a systematic analysis of $H_{(s)}\bar{T}_{(s)}$ and $H_{(s)}T_{(s)}$ configurations through the implementation of the one-boson-exchange (OBE) model~\cite{Ogawa:1967iv,DeSwart:1993te,Machleidt:1987hj,Machleidt:1989tm,DeTourreil:1975gz, Holinde:1976mkn, Nagels:1977ze}, a well-validated approach in characterizing hadronic interactions~\cite{Chen:2022asf,Chen:2016qju,Liu:2019zoy}. 
Within this framework, the $H_{(s)}$ and $T_{(s)}$ doublets correspond to $(B_{(s)},B_{(s)}^*)$ and $(B_{1(s1)},B^{*}_{2(s2)})$, respectively. 
The Gaussian expansion method (GEM)~\cite{Hiyama:2003cu,Hiyama:2018ivm} is employed to solve the Schrödinger equation to systematically identify the weakly bound states. 
Through parametric variation of the cutoff $\Lambda$ around $1000~\rm{MeV}$, distinct bound states emerge in the following channels: $I(J^{PC})=0(1^{-\pm})$ $B\bar{B}_{1}$, $I(J^{PC})=0(2^{-\pm})$ $B\bar{B}_{2}^{*}$, $I(J^{PC})=0(1^{-\pm})$ $B^*\bar{B}_{1}$, $I(J^{PC})=0(2^{-\pm})$ $B^*\bar{B}_{2}^{*}$,  $I(J^{PC})=0(1^{-})$ $B^*B_1$, and $I(J^{PC})=0(2^{-})$ $B^{*}B^{*}_2$. 
Conversely, no bound state is obtained in the bottom-strange sectors. 
These theoretical predictions are anticipated to provide important information for forthcoming experimental searches by the LHCb and Belle II Collaborations.

This paper is organized as follows. 
In Sec.~\ref{sec2}, we present the detailed derivation of the $S$- and $D$-wave interactions between the bottom(-strange) mesons in $H_{(s)}$ and $T_{(s)}$ doublets within the OBE model. 
The predicted bound states and relevant discussions are presented in Sec.~\ref{sec3}. 
Finally, we present a brief summary in Sec.~\ref{sec4}.
\section{FORMALISM}\label{sec2}
{In the present work, the OBE model is applied to obtain the effective potentials for the $H_{(s)}\bar{T}_{(s)}$ and  $H_{(s)}T_{(s)}$ systems, which can describe the meson-meson interactions by exchanging light mesons well. 
Within the OBE model, the long-range interaction 
is dominated by the potential arising from the one pion  exchange, while the short-range  interaction  mainly  comes from the vector meson  exchanges.}
The effective Lagrangian was constructed through the incorporation of heavy quark symmetry and chiral symmetry, following established theoretical methodologies~\cite{Yan:1992gz,Wise:1992hn,Burdman:1992gh, Casalbuoni:1996pg}
\begin{eqnarray}
{\mathcal L}&=&g_{\sigma}\left\langle H^{(Q)}_a\sigma\bar{H}^{(Q)}_a\right\rangle+g_{\sigma}\left\langle \bar{H}^{(\bar{Q})}_a\sigma H^{(\bar{Q})}_a\right\rangle\nonumber\\
&&+ig\left\langle H^{(Q)}_b{\mathcal A}\!\!\!\slash_{ba}\gamma_5\bar{H}^{\,({Q})}_a\right\rangle+ig\left\langle \bar{H}^{(\bar{Q})}_a{\mathcal A}\!\!\!\slash_{ab}\gamma_5 H^{\,(\bar{Q})}_b\right\rangle\nonumber\\
&&+\left\langle iH^{(Q)}_b\left(\beta v^{\mu}({\mathcal V}_{\mu}-\rho_{\mu})+\lambda \sigma^{\mu\nu}F_{\mu\nu}(\rho)\right)_{ba}\bar{H}^{\,(Q)}_a\right\rangle\nonumber\\
&&-\left\langle i\bar{H}^{(\bar{Q})}_a\left(\beta v^{\mu}({\mathcal V}_{\mu}-\rho_{\mu})-\lambda \sigma^{\mu\nu}F_{\mu\nu}(\rho)\right)_{ab}H^{\,(\bar{Q})}_b\right\rangle,
\end{eqnarray}
where the  $a$ and $b$ are flavor indices and  $v^{\mu}=(1, \bf{0})$ denotes the four-velocity, the explicit expressions for the vector current $\mathcal{V}_{\mu}$, axial-vector current ${A}_{\mu}$, and field strength tensor $F_{\mu\nu}(\rho)$ are defined as
\begin{eqnarray}
       \mathcal{V}_{\mu} &=& \frac{1}{2}(\xi^{\dag}\partial_{\mu}\xi+\xi\partial_{\mu}\xi^{\dag}),\\
       A_{\mu} &=& \frac{1}{2}(\xi^{\dag}\partial_{\mu}\xi-\xi\partial_{\mu}\xi^{\dag}),\\
       F_{\mu\nu}(\rho) &=& \partial_{\mu}\rho_{\nu}-\partial_{\nu}\rho_{\mu}+[\rho_{\mu},\rho_{\nu}].
  \end{eqnarray}
Here, $\xi=\text{exp}(i{\mathbb{P}}/f_{\pi})$ and {$\rho_{\mu}=ig_{V}{\mathbb{V}}/\sqrt{2}$}. The $f_{\pi}=132$ MeV
is the pion-decay constant~\cite{Wang:2021yld,Chen:2022svh,Song:2025yut}, and then $g_V=m_{\rho}/f_{\pi}=5.8$~\cite{Bando:1987br}. The $\mathbb{P}$ and $\mathbb{V}$ stand for the matrices of light pseudoscalar and vector mesons, respectively,  
\begin{eqnarray}
\left.\begin{array}{c} {\mathbb{P}} = {\left(\begin{array}{ccc}
       \frac{\pi^0}{\sqrt{2}}+\frac{\eta}{\sqrt{6}} &\pi^+ &K^+\\
       \pi^-       &-\frac{\pi^0}{\sqrt{2}}+\frac{\eta}{\sqrt{6}} &K^0\\
       K^-         &\bar K^0   &-\sqrt{\frac{2}{3}} \eta     \end{array}\right)},\\
{\mathbb{V}} = {\left(\begin{array}{ccc}
       \frac{\rho^0}{\sqrt{2}}+\frac{\omega}{\sqrt{2}} &\rho^+ &K^{*+}\\
       \rho^-       &-\frac{\rho^0}{\sqrt{2}}+\frac{\omega}{\sqrt{2}} &K^{*0}\\
       K^{*-}         &\bar K^{*0}   & \phi     \end{array}\right)}.
\end{array}\right.
\end{eqnarray}
The $H^{(Q)}_a$, $H^{(\bar Q)}_a$, $\bar H^{(Q)}_a$, and $\bar H^{(\bar Q)}_a$ represent the fields of heavy-light mesons   and can be written as 
\begin{eqnarray}
H^{(Q)}_a&=&(1+{v}\!\!\!\slash)(\mathcal{P}^{*(Q)\mu}_a\gamma_{\mu}-\mathcal{P}^{(Q)}_a\gamma_5)/2,\\
H^{(\bar{Q})}_a &=& (\bar{\mathcal{P}}^{*(\bar{Q})\mu}_a\gamma_{\mu}-\bar{\mathcal{P}}^{(\bar{Q})}_a\gamma_5)
(1-{v}\!\!\!\slash)/{2},\\
\bar{H}&=&\gamma_0H^{\dagger}\gamma_0.
\end{eqnarray}
The expanded effective Lagrangians for describing the interactions between the heavy-light mesons can be found in Refs.~\cite{Wang:2020dya, Wang:2019nwt, Chen:2018pzd}.

With the effective Lagrangian, one can derive the effective OBE potentials in the coordinate space for the $H_{(s)}\bar{T}_{(s)}$ and $H_{(s)}T_{(s)}$ systems~\cite{Wang:2020dya, Wang:2019nwt, Wang:2019aoc, Wang:2020bjt, Wang:2021hql, Chen:2018pzd, Wang:2021yld}. 
That is to say, the effective potentials in the momentum space $\mathcal{V}^{h_1h_2\to h_3h_4}_E(\bm{q})$ can be obtained based on the Breit approximation~\cite{Breit:1929zz,Breit:1930zza} and the non-relativistic normalization. 
\begin{eqnarray}\label{breit}
\mathcal{V}_E^{h_1h_2\to h_3h_4}(\bm{q})=-\frac{\mathcal{M}(h_1h_2\to h_3h_4)} {\sqrt{2m_{h_1}2m_{h_2}2m_{h_3}2m_{h_4}}}.
\end{eqnarray}
Subsequently, the effective potential in coordinate space is obtained from its momentum-space counterpart via Fourier transformation. In the heavy quark limit, $Q_{\mathbf 3}\bar{q}_{\bar{\mathbf 3}}$ mesons and $Q_{\mathbf 3} [qq]_{\bar{\mathbf 3}}$ baryons possess antiquark-diquark symmetry. Therefore, the $Q\bar{q}\, q^\prime \bar{Q}$ (tetraquark) and $Q[qq]\, q^\prime \bar{Q}$ (pentaquark) systems can thus be analyzed within the same theoretical framework~\cite{Wang:2025sic}. We construct the potential function following Refs.~\cite{Wang:2020dya, Wang:2019nwt, Wang:2019aoc, Wang:2020bjt, Wang:2021hql, Chen:2018pzd, Wang:2021yld}, with the specific formulas detailed in Appendix~\ref{app:1}.

Following the derivation of effective potentials, the coupled-channel Schr\"odinger equation governing the $H_{(s)}\bar{T}_{(s)}$ and $H_{(s)}T_{(s)}$ systems is numerically solved, incorporating $S$-$D$ wave coupling effects. Meanwhile, the Gaussian expansion method transforms the Schr\"odinger equation into a generalized eigenvalue formulation through wave function decomposition onto Gaussian basis functions. 
Through the implementation of the variational principle, the masses and corresponding wave functions for the exotic hadronic systems are systematically extracted from the numerical solutions of the generalized eigenvalue formulation.

\section{Results and discussions}\label{sec3}

\begin{figure*}
	\centering
	\includegraphics[scale=0.60]{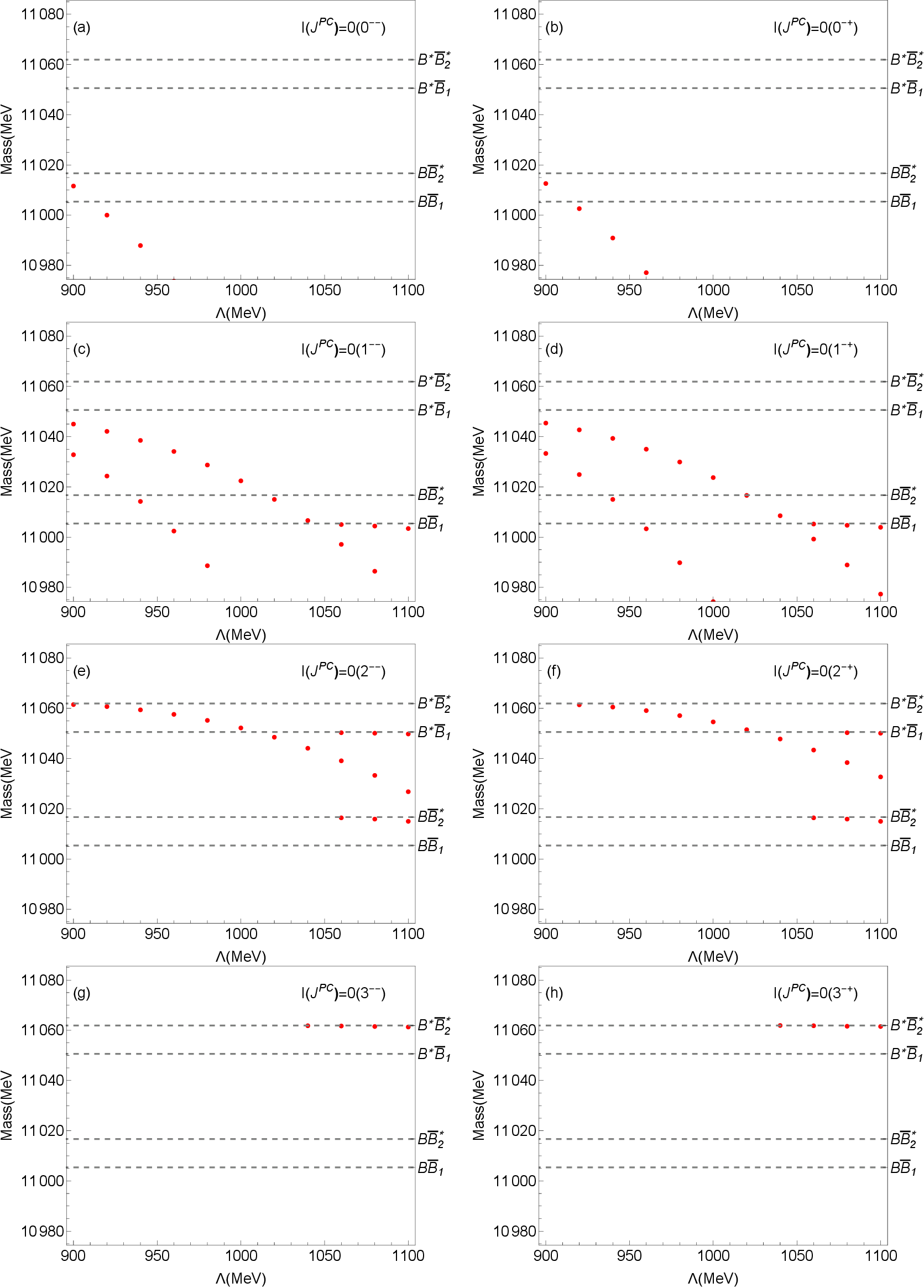}
	\caption{The $\Lambda$  dependence for the $H\bar{T}$ systems. The red solid dots stand for the bound states. }
	\label{bb}
\end{figure*}

Subsequent to the implementation of the aforementioned computational procedures, 
the solutions for bound states can be systematically examined 
by numerically solving the Schr\"odinger equation.
It is important to note that the cutoff parameter, denoted by $\Lambda$, is the only adjustable variable, and it reflects the size of the constituent heavy hadrons.
Our previous theoretical framework, as referenced in~\cite{Song:2024ngu} provided a practical methodology 
for investigating hidden-bottom molecular tetraquark configurations $B^{(*)}_{(s)}\bar{B}^{(*)}_{(s)}$, it was suggested that within the phenomenological interval $\Lambda$ in the range of $950 \sim 1200~\rm{MeV}$, 
the OBE formalism can reproduce the mass spectra of the exotic $Z_b(10610)$ and $Z_b(10650)$ resonances.
Consequently, we chose a characteristic scale of $\Lambda$ around $1000~\rm{MeV}$.
Moreover, numerical solutions that yield a root-mean-square (RMS) radius of less than $1 ~\rm{fm}$ are {discarded}. {The reason for this exclusion is that such numerical results lie outside the regime of validity of the OBE model employed in this work. In this model, hadrons are treated as point-like particles, and they are characterized by a cutoff corresponding to a length scale. That is, the OBE model can only predict the existence of loosely bound or resonant molecules rather than compact exotic states.} Consequently, {this physical criterion is} enforced in the present work to identify and discard {such improper solutions}, thereby ensuring the reliability of our predictions.
Initially, our computations focus on the $H\bar{T}$ systems, specifically the $B\bar{B}_1$, 
$B\bar{B}^*_2$, $B^*\bar{B}_1$, and $B^*\bar{B}^*_2$ configurations.
This study then extends to the $HT$ systems, followed by theoretical estimates for bottom-strange combinations.
The entirety of the corresponding results are comprehensively summarized in Tables ~\ref{sr1}-\ref{sr12} 
and visually  demonstrated in Fig.~\ref{bb}.

\subsection{The $B\bar{B}_1$, $B\bar{B}^*_2$, $B^*\bar{B}_1$, and $B^*\bar{B}^*_2$ systems}
For $B\bar{B}_{1}$ channel, there exist spin-parities $I(J^{PC})=0(1^{--})$, $0(1^{-+})$, $1(1^{--})$, and $1(1^{-+})$. According to our calculations, we find loosely bound states for the $0(1^{--})$ and $0(1^{-+})$ configurations, where the cutoff value is consistent with the empirical value for deuteron well and can be regarded as good molecular states~\cite{Tornqvist:1993ng,Tornqvist:1993vu,Wang:2019nwt,Chen:2017jjn,Chen:2016qju,Liu:2019zoy}. It can be seen that the mass of $\Upsilon(10753)$ is much lower than the $B \bar B_{s1}$ threshold, which can not be interpreted as the $0(1^{--})$ $B \bar B_{s1}$ molecular state in our calculations. Moreover, if we discard the short-range $\sigma$ exchange contribution, our conclusion remains, which suggests that the $\omega$ and $\rho$ exchange potentials are mainly responsible for the molecular formation. However, for the higher isospin ones, we can not obtain any bound state since the flavor factors in the potentials for $I=0$ and $I=1$ configurations are quite different in the  $B\bar{B}_1$ channel.

\renewcommand\tabcolsep{0.50cm}
\renewcommand{\arraystretch}{1.50}
\begin{table}[!htbp]
\caption{{Numerical solutions for bound states with $S-D$ wave mixing in the $B\bar{B}_{1}$ system. The units for the cutoff $\Lambda$, binding energy $E$, and RMS radius $r_{\mathrm{RMS}}$ are $\mathrm{GeV}$, $\mathrm{MeV}$, and $\mathrm{fm}$, respectively. A "$\times$" denotes the absence of a bound state for a cutoff scanned from 0.8 to 3.0~$\mathrm{GeV}$. Probabilities for the dominant channels are shown in bold, and the most promising molecular candidates are marked with $\checkmark$.}}\label{sr1}

\begin{tabular}{c|cccl}\toprule[1.0pt]\toprule[1.0pt]
$I(J^{PC})$&$\Lambda$      &$E$&$r_{\rm RMS}$&$P({}^3\mathbb{S}_{1}/{}^3\mathbb{D}_{1})$  \\\midrule[1.0pt]
\multirow{3}{*}{{$ 0(1^{--})\checkmark$}}          &1.05&$0.15$&3.80&\textbf{99.98}/0.02  \\
                                    &1.10&$2.01$&1.64&\textbf{99.94}/0.06  \\
                                    &1.16&$6.67$&1.03&\textbf{99.92}/0.08 \\
\multirow{1}{*}{$1(1^{--})$}         &$\times$&$\times$&$\times$ &$\times$\\                            
                                    \midrule[1.0pt]
\multirow{3}{*}{{$0(1^{-+})\checkmark$}}         &1.06&$0.24$&3.55&\textbf{99.97}/0.03  \\
                                    &1.12&$2.52$&1.54&\textbf{99.93}/0.07  \\
                                    &1.18&$6.97$&1.04&\textbf{99.90}/0.10 \\
\multirow{1}{*}{$1(1^{-+})$}        &$\times$&$\times$&$\times$ &$\times$\\                              
\bottomrule[1pt]\bottomrule[1pt]
\end{tabular}

\end{table}


In the $B\bar{B}^*_{2}$ channel, we consider both $I(J^{P}) = 0(2^{-\pm})$ and $1(2^{-\pm})$ states. In our calculations, we neglect the effects of higher-order terms and discard the $C$-parity state for $B\bar{B}^*_{2}$ channel. For the isospin $I = 0$ $B\bar{B}^*_{2}$ channel, we enlarge the cutoff from $1060\sim1170~\rm MeV$, a bound state with binding energies ranging from $0.27$ to $6.71~\rm MeV$ is obtained, where the root-mean-square (RMS) radius $r_{RMS}$ is in a range of $1.04\sim3.41~\rm fm$. This result suggests that the $I(J^{P}) = 0(2^{-\pm})$ $B\bar{B}^*_{2}$ channels could potentially form a promising molecular state, with the corresponding results shown in Table~\ref{sr2}. On the other hand, for the isospin $I = 1$ $B\bar{B}^*_{2}$ channel, our calculations indicate that no bound state is formed. This highlights the importance of stronger interactions in the $I = 0$ states, while the weaker interactions in the $I = 1$ states hinder the formation of a molecular state.

\renewcommand\tabcolsep{0.50cm}
\renewcommand{\arraystretch}{1.50}
\begin{table}[!htbp]
\caption{Numerical solutions for the bound states with the $S-D$ wave mixing in the $B\bar{B}_{2}^*$ system. Conventions are the same as those in Table~\ref{sr1}.}\label{sr2}
\begin{tabular}{c|cccl}\toprule[1.0pt]\toprule[1.0pt]

$I(J^{P})$&$\Lambda$      &$E$&$r_{\rm RMS}$&$P({}^5\mathbb{S}_{2}/{}^5\mathbb{D}_{2})$  \\\midrule[1.0pt]
\multirow{3}{*}{{$0(2^{\pm}) \checkmark$}}         &1.06&$0.27$&3.41&\textbf{100}/0.00  \\
                                    &1.11&$2.21$&1.60&\textbf{100}/0.00\\
                                    &1.17&$6.71$&1.04&\textbf{100}/0.00 \\
\multirow{1}{*}{$1(2^{-\pm})$}        &$\times$&$\times$&$\times$ &$\times$\\                           
\bottomrule[1pt]\bottomrule[1pt]
\end{tabular}
\end{table}


Compared to the above $B\bar{B}_1$ and $B\bar{B}^*_{2}$ channels, the $B^*\bar{B}_1$ channel exhibits a more complex combination of quantum states and significant $S$-$D$ wave mixing effects. As shown in Table~\ref{sr3}, the following states exist $I(J^{PC})=0(0^{--})$, $1(0^{--})$, $0(0^{-+})$, $1(0^{-+})$, $0(1^{--})$, $0(1^{-+})$, $1(1^{--})$, $1(1^{-+})$, $0(2^{--})$, $0(2^{-+})$, $1(2^{--})$, and $1(2^{-+})$ states. According to our estimations, we predict two loosely bound states in the $I(J^{PC})=0(1^{--})$ and $0(1^{-+})$ states in the cutoff range $800\sim900$ $\rm MeV$, where the $r_{RMS}$ varies from $4\sim1$ $\rm fm$ and they are regarded as the most promising molecular states. For the $I(J^{PC})=0(0^{--})$ and $0(0^{-+})$ configurations, one can obtain another two bound states when the cutoff is around $1160$ $\rm MeV$. It should be mentioned that we can also obtain two bound states for $0(2^{--})$ and $0(2^{-+})$ states that are insensitive to the cutoff parameter. In contrast to the $I=0$ case, the $I=1$ states require higher cutoff parameters to form bound states, further demonstrating that higher isospin quantum numbers have a suppressive effect on the existence of bound states. This observation is consistent with our previous theoretical predictions for molecular states, which suggest that lower isospin channels are more likely to form molecular states~\cite{Song:2024ngu,Lu:2024dtb}.

\renewcommand\tabcolsep{0.31cm}
\renewcommand{\arraystretch}{1.50}
\begin{table}[!htbp]
\caption{Numerical solutions for the bound states with the $S-D$ wave mixing in the $B^*\bar{B}_{1}$ system. Conventions are the same as those in Table~\ref{sr1}.}\label{sr3}
\begin{tabular}{c|cccl}\toprule[1.0pt]\toprule[1.0pt]
$I(J^{PC})$&$\Lambda$      &$E$&$r_{\rm RMS}$&$P({}^1\mathbb{S}_{0}/{}^5\mathbb{D}_{0})$  \\\midrule[1.0pt]
\multirow{3}{*}{$0(0^{--})$}         &1.15&$0.09$&3.95&\textbf{97.84}/2.13  \\
                                    &1.18&$1.88$&1.73&\textbf{96.61}/3.39 \\
                                    &1.22&$7.34$&1.07&\textbf{96.67}/3.33 \\
\multirow{1}{*}{$1(0^{--})$}         &$\times$&$\times$&$\times$ &$\times$\\                                
                                   \midrule[1.0pt]
\multirow{3}{*}{$0(0^{-+})$}         &1.16&$0.32$&3.21&\textbf{97.58}/2.42  \\
                                    &1.19&$2.44$&1.57&\textbf{96.80}/3.20  \\
                                    &1.23&$8.29$&1.03&\textbf{96.99}/3.01 \\
\multirow{1}{*}{$1(0^{-+})$}         &$\times$&$\times$&$\times$ &$\times$\\                            
                                    \midrule[1.0pt]
$I(J^{PC})$&$\Lambda$ &$E$&$r_{\rm RMS}$&$P({}^3\mathbb{S}_{1}/{}^3\mathbb{D}_{1}/{}^5\mathbb{D}_{1})$ \\
\multirow{3}{*}{{$0(1^{--})\checkmark $}}         &0.81&$0.18$&3.62&\textbf{98.10}/1.90/0.00  \\
                                    &0.86&$1.98$&1.65&\textbf{97.18}/2.81/0.00  \\
                                    &0.91&$6.96$&1.01&\textbf{97.25}/2.75/0.00  \\
\multirow{1}{*}{$1(1^{--})$}          &$\times$&$\times$&$\times$ &$\times$\\                           
                                    \midrule[1.0pt]             
\multirow{3}{*}{{$(1^{-+}) \checkmark$}}         &0.81&$0.12$&3.88&\textbf{98.25}/1.75/0.00  \\
                                    &0.86&$1.74$&1.74&\textbf{97.28}/2.72/0.00  \\
                                    &0.91&$6.43$&1.04&\textbf{97.33}/2.67/0.00  \\
\multirow{1}{*}{$1(1^{-+})$}         &$\times$&$\times$&$\times$ &$\times$\\                            
                                    \midrule[1.0pt]                
$I(J^{PC})$&$\Lambda$&$E$&$r_{\rm RMS}$&$P({}^5\mathbb{S}_{2}/{}^1\mathbb{D}_{2}/{}^3\mathbb{D}_{2}/{}^5\mathbb{D}_{2})$ \\   
\multirow{3}{*}{$0(2^{--})$}         &1.05&$0.22$&3.89&\textbf{95.45}/0.25/0.00/4.30  \\
                                    &1.35&$7.51$&1.24&\textbf{89.44}/0.72/0.00/9.84\\
                                    &1.50&$13.75$&1.02&\textbf{88.79}/1.32/0.00/9.89\\
\multirow{3}{*}{$1(2^{--})$}         &2.40&$0.08$&3.99&\textbf{98.86}/0.18/0.00/0.96  \\
                                    &2.70&$2.14$&1.46&\textbf{97.15}/0.45/0.00/2.40\\      
                                    &2.95&$4.86$&1.02&\textbf{96.45}/0.57/0.00/2.99\\  
                                    \midrule[1.0pt]                          
\multirow{3}{*}{$0(2^{-+})$}         &1.07&$0.24$&3.82&\textbf{95.51}/0.26/0.00/4.23  \\
                                    &1.31&$4.96$&1.42&\textbf{90.73}/0.61/0.00/8.66  \\
                                    &1.55&$13.63$&1.02&\textbf{89.82}/1.70/0.00/8.48  \\
\multirow{3}{*}{$1(2^{-+})$}        &1.90&$0.28$&3.24&\textbf{98.23}/0.29/0.00/1.48  \\
                                    &2.10&$2.20$&1.46&\textbf{96.62}/0.55/0.00/2.83 \\
                                    &2.25&$5.09$&1.02&\textbf{95.78}/0.69/0.00/3.53 \\                            
\bottomrule[1pt]\bottomrule[1pt]
\end{tabular}
\end{table}


The configurations for the $B^*\bar{B}_2^*$ system include $I(J^{PC})=0(0^{-\pm})$, $1(0^{-\pm})$, $0(1^{-\pm})$, $1(1^{-\pm})$, $0(2^{-\pm})$, $1(2^{-\pm})$, $0(3^{-\pm})$, and $1(3^{-\pm})$ states. Theoretical analysis in Table~\ref{sr4} reveals that the bound-state solutions emerge in the isospin-singlet channels spanning angular momentum configurations from $J^{PC}=0^{-\pm}$ to $3^{-\pm}$, with the characteristic cutoff parameter around $1000~\rm MeV$.

\renewcommand\tabcolsep{0.31cm}
\renewcommand{\arraystretch}{1.50}
\begin{table}[!htbp]
\caption{Numerical solutions for the bound states with the $S-D$ wave mixing in the $B^*\bar{B}_{2}^*$ system. Conventions are the same as those in Table~\ref{sr1}.}\label{sr4}
\begin{tabular}{c|cccl}\toprule[1.0pt]\toprule[1.0pt]
$I(J^{PC})$&$\Lambda$&$E$&$r_{\rm RMS}$&$P({}^3\mathbb{S}_{1}/{}^3\mathbb{D}_{1}/{}^5\mathbb{D}_{1}/{}^7\mathbb{D}_{1})$ \\
\multirow{1}{*}{$0(1^{--})$}         &0.78&$4.87$&1.12&\textbf{99.26}/0.30/0.00/0.44  \\
                                     &0.80&$6.50$&1.00&\textbf{99.28}/0.29/0.00/0.43  \\
                                     &0.82&$8.87$&0.89&\textbf{99.32}/0.28/0.00/0.40  \\
\multirow{1}{*}{$1(1^{--})$}          &$\times$&$\times$&$\times$ &$\times$\\                            
                                    \midrule[1.0pt]             
\multirow{1}{*}{$0(1^{-+})$}         &0.78&$4.79$&1.12&\textbf{99.07}/0.46/0.00/0.47 \\
                                     &0.80&$6.39$&1.01&\textbf{99.06}/0.48/0.00/0.46 \\
                                     &0.82&$8.73$&0.90&\textbf{99.08}/0.48/0.00/0.44 \\
\multirow{1}{*}{$1(1^{-+})$}         &$\times$&$\times$&$\times$ &$\times$\\                             
                                    \midrule[1.0pt]                
$I(J^{PC})$&$\Lambda$&$E$&$r_{\rm RMS}$&$P({}^5\mathbb{S}_{2}/{}^3\mathbb{D}_{2}/{}^5\mathbb{D}_{2}/{}^7\mathbb{D}_{2})$ \\   
\multirow{3}{*}{{$0(2^{--}) \checkmark$}}         &0.89&$0.12$&3.87&\textbf{98.35}/0.00/1.65/0.00  \\
                                    &0.94&$2.45$&1.51&\textbf{97.23}/0.00/2.77/0.00\\
                                    &0.98&$6.70$&1.02&\textbf{97.29}/0.00/2.71/0.00  \\ 
\multirow{1}{*}{$1(2^{--})$}        &$\times$&$\times$&$\times$ &$\times$\\ 
                                    \midrule[1.0pt]                          
\multirow{3}{*}{{$0(2^{-+}) \checkmark$}}         &0.91&0.23&$3.48$&\textbf{98.50}/0.00/1.50/0.00  \\
                                    &0.95&$2.08$&1.62&\textbf{97.91}/0.00/2.09/0.00 \\
                                    &0.99&$5.98$&1.07&\textbf{98.04}/0.00/1.96/0.00  \\
\multirow{1}{*}{$1(2^{-+})$}       &$\times$&$\times$&$\times$ &$\times$\\ 
                                    \midrule[1.0pt]                
$I(J^{PC})$&$\Lambda$&$E$&$r_{\rm RMS}$&$P({}^7\mathbb{S}_{3}/{}^3\mathbb{D}_{3}/{}^5\mathbb{D}_{3}/{}^7\mathbb{D}_{3})$ \\   
\multirow{3}{*}{$0(3^{--})$}        
                                    &1.07&$0.27$&3.76&\textbf{94.46}/0.17/0.00/5.37\\ 
                                    
                                      &1.32&$5.39$&1.40&\textbf{87.55}/0.34/0.00/12.11  \\
                                      
                                        &1.57&$14.75$&1.02&\textbf{85.99}/0.58/0.00/13.43  \\

\multirow{3}{*}{$1(3^{--})$}       &1.52&$0.14$&3.68&\textbf{99.39}/0.11/0.00/0.50\\ 
                                   &1.64&$1.55$&1.65&\textbf{98.98}/0.21/0.00/0.80 \\
                                   &1.76&$4.62$&1.02&\textbf{98.86}/0.26/0.00/0.88  \\
                                   
                                    \midrule[1.0pt]                          
\multirow{3}{*}{$0(3^{-+})$}         &1.08&$0.26$&3.78&\textbf{95.21}/0.21/0.00/4.58  \\
                                    &1.35&$5.64$&1.39&\textbf{90.17}/0.57/0.00/9.26  \\                     &1.62&$15.69$&1.00&\textbf{89.55}/2.02/0.00/8.43 \\ 
                                    
\multirow{3}{*}{$1(3^{-+})$}         &2.35&$0.20$&3.54&\textbf{98.06}/0.11/0.00/1.83  \\
                                    &2.65&$1.74$&1.66&\textbf{95.86}/0.21/0.00/3.93 \\                     &2.95&$5.21$&1.05&\textbf{94.02}/0.29/0.00/5.69 \\ 
                                                               
\bottomrule[1pt]\bottomrule[1pt]
\end{tabular}
\end{table}

In all channels, the $S$-wave components dominate, indicating that these molecular states are primarily governed by their lowest orbital angular momentum ($L=0$) configuration in their ground state. Additionally, we observe a slight rise in the $D$-wave contribution as the total angular momentum $J$ increases. For example, in the $0(3^{--})$ and $0(3^{-+})$ states, the $D$-wave contribution can reach a few percent. This reflects the significance of tensor forces in these channels; as $J$ increases, tensor forces promote the coupling between the $S$ and $D$ waves, and thereby increase the probabilities of $D$-wave components.


\subsection{The $BB_1$, $BB^*_2$, $B^*B_1$, and $B^*B^*_2$ systems}

In this subsection, we examine the solutions for  bound states with the $S$-$D$ wave mixing for the $BB_{1}$, $BB^*_{2}$, $B^*B_{1}$, and $B^*B^*_{2}$ systems. Relevant numerical results are listed in Tables~\ref{sr5} to \ref{sr8}. For the $BB_1$ system, a bound state with $I(J^P)=1^-$ is found with a cutoff in the range $1210\sim 1390$ $\rm MeV$ and the corresponding $r_{RMS}$ varies from $4\sim 1~\rm fm$, while we cannot obtain any bound state for the $I=1$ configuration. For the $BB^*_{2}$ system, the situation also shows similar behaviors to the $BB_{1}$ case.

\renewcommand\tabcolsep{0.45cm}
\renewcommand{\arraystretch}{1.50}
\begin{table}[!htbp]
\caption{Numerical solutions for the bound states with the $S-D$ wave mixing in the $B{B}_{1}$ system. Conventions are the same as those in Table~\ref{sr1}.}\label{sr5}
\begin{tabular}{c|cccl}\toprule[1.0pt]\toprule[1.0pt]
$I(J^{P})$&$\Lambda$      &$E$&$r_{\rm RMS}$&$P({}^3\mathbb{S}_{1}/{}^3\mathbb{D}_{1})$  \\
\multirow{3}{*}{$0(1^{-})$}         &1.21&$0.20$&3.57&\textbf{99.96}/0.04  \\
                                    &1.30&$2.31$&1.62&\textbf{99.88}/0.12  \\
                                    &1.39&$6.34$&1.01&\textbf{99.82}/0.17 \\
                                    
\multirow{1}{*}{$1(1^{-})$}         &$\times$&$\times$&$\times$ &$\times$\\                            
                                      
\bottomrule[1pt]\bottomrule[1pt]
\end{tabular}
\end{table}
\renewcommand\tabcolsep{0.45cm}
\renewcommand{\arraystretch}{1.50}
\begin{table}[!htbp]
\caption{Numerical solutions for the bound states with the $S-D$ wave mixing in the $B{B}_{2}^*$ system. Conventions are the same as those in Table~\ref{sr1}.}\label{sr6}
\begin{tabular}{c|cccl}\toprule[1.0pt]\toprule[1.0pt]

$I(J^{P})$&$\Lambda$      &$E$&$r_{\rm RMS}$&$P({}^5\mathbb{S}_{2}/{}^5\mathbb{D}_{2})$  \\\midrule[1.0pt]
\multirow{3}{*}{$0(2^{-})$}         &1.23&$0.15$&3.80&\textbf{100}/0.00  \\
                                    &1.34&$2.31$&1.34&\textbf{100}/0.00  \\
                                    &1.45&$6.51$&1.02&\textbf{100}/0.00\\
\multirow{1}{*}{$1(2^{-})$} &$\times$&$\times$&$\times$ &$\times$\\
                                   
\bottomrule[1pt]\bottomrule[1pt]
\end{tabular}
\end{table}
\renewcommand\tabcolsep{0.31cm}
\renewcommand{\arraystretch}{1.50}
\begin{table}[!htbp]
\caption{Numerical solutions for the bound states with the $S-D$ wave mixing in the $B^*{B}_{1}$ system. Conventions are the same as those in Table~\ref{sr1}.}\label{sr7}
\begin{tabular}{c|cccl}\toprule[1.0pt]\toprule[1.0pt]
$I(J^{P})$&$\Lambda$      &$E$&$r_{\rm RMS}$&$P({}^1\mathbb{S}_{0}/{}^5\mathbb{D}_{0})$  \\\midrule[1.0pt]
\multirow{3}{*}{$0(0^{-})$}         &1.32&$0.16$&3.64&\textbf{97.55}/2.45 \\
                                    &1.37&$2.39$&1.49&\textbf{96.20}/3.80 \\
                                    &1.42&$7.05$&1.05&\textbf{96.14}/3.86 \\
\multirow{1}{*}{$1(0^{-})$}         &$\times$&$\times$&$\times$ &$\times$\\                                
                                 
                                    \midrule[1.0pt]
$I(J^{P})$&$\Lambda$ &$E$&$r_{\rm RMS}$&$P({}^3\mathbb{S}_{1}/{}^3\mathbb{D}_{1}/{}^5\mathbb{D}_{1})$ \\
\multirow{3}{*}{{$0(1^{-}) \checkmark$}}         &0.84&$0.31$&3.26&\textbf{97.88}/2.11/0.00  \\
                                    &0.90&$2.19$&1.58&\textbf{97.24}/2.76/0.00  \\
                                    &0.96&$6.52$&1.03&\textbf{97.26}/2.74/0.00  \\
\multirow{1}{*}{$1(1^{-})$}          &$\times$&$\times$&$\times$ &$\times$\\                           
                                    \midrule[1.0pt]             
              
$I(J^{P})$&$\Lambda$&$E$&$r_{\rm RMS}$&$P({}^5\mathbb{S}_{2}/{}^1\mathbb{D}_{2}/{}^3\mathbb{D}_{2}/{}^5\mathbb{D}_{2})$ \\   
\multirow{3}{*}{{$0(2^{-}) \checkmark$}}         &1.25&$0.44$&3.28&\textbf{94.27}/0.31/0.00/5.42  \\
                                    &1.60&$4.72$&1.41&\textbf{89.03}/0.59/0.00/10.38\\
                                    &1.95&$12.12$&1.08&\textbf{86.98}/1.06/0.00/11.96\\
\multirow{2}{*}{$1(2^{-})$}         &1.80&$0.34$&2.4&\textbf{99.50}/0.08/0.00/0.42  \\
                                    &1.81&$2.08$&1.01&\textbf{99.57}/0.07/0.00/0.36\\
                                                        
\bottomrule[1pt]\bottomrule[1pt]
\end{tabular}
\end{table}
\renewcommand\tabcolsep{0.31cm}
\renewcommand{\arraystretch}{1.50}
\begin{table}[!htbp]
\caption{Numerical solutions for the bound states with the $S-D$ wave mixing in the $B^*B^*_{2}$  system. Conventions are the same as those in Table~\ref{sr1}.}\label{sr8}
\begin{tabular}{c|cccl}\toprule[1.0pt]\toprule[1.0pt]
$I(J^{P})$&$\Lambda$&$E$&$r_{\rm RMS}$&$P({}^3\mathbb{S}_{1}/{}^3\mathbb{D}_{1}/{}^5\mathbb{D}_{1}/{}^7\mathbb{D}_{1})$ \\
\multirow{3}{*}{$0(1^{-})$}         &0.80&$5.29$&1.08&\textbf{99.29}/0.29/0.00/0.42  \\
                                    &0.81&$6.06$&1.02&\textbf{99.30}/0.28/0.00/0.42  \\
                                    &0.82&$6.94$&0.97&\textbf{99.31}/0.27/0.00/0.41  \\
                                    
\multirow{1}{*}{$1(1^{-})$}          &$\times$&$\times$&$\times$ &$\times$\\                            
                                    \midrule[1.0pt]             
$I(J^{P})$&$\Lambda$&$E$&$r_{\rm RMS}$&$P({}^5\mathbb{S}_{2}/{}^3\mathbb{D}_{2}/{}^5\mathbb{D}_{2}/{}^7\mathbb{D}_{2})$ \\   
\multirow{3}{*}{{$0(2^{-}) \checkmark$}}         
                                    &0.93&$0.12$&3.89&\textbf{98.32}/0.00/1.68/0.00\\
                                    &1.00&$2.37$&1.52&\textbf{97.02}/0.00/2.98/0.00\\
                                    &1.06&$6.78$&1.01&\textbf{96.92}/0.00/3.08/0.00 \\
\multirow{1}{*}{$1(2^{-})$}        &$\times$&$\times$&$\times$ &$\times$\\ 
                                    \midrule[1.0pt]                          
$I(J^{P})$&$\Lambda$&$E$&$r_{\rm RMS}$&$P({}^7\mathbb{S}_{3}/{}^3\mathbb{D}_{3}/{}^5\mathbb{D}_{3}/{}^7\mathbb{D}_{3})$ \\   
\multirow{3}{*}{$0(3^{-})$}        
                                     &1.18&$0.23$&3.86&\textbf{94.33}/0.17/0.00/5.50\\ 
                                   &1.51&$4.63$&1.43&\textbf{85.66}/0.31/0.00/14.04 \\
                                   &1.84&$13.49$&1.00&\textbf{81.89}/0.37/0.00/17.74  \\
                                
\multirow{3}{*}{$1(3^{-})$}       &1.46&$0.37$&2.53&\textbf{99.70}/0.07/0.00/0.23\\ 
                                   &1.47&$1.31$&1.44&\textbf{99.72}/0.07/0.00/0.21 \\
                                   &1.48&$2.59$&1.04&\textbf{99.77}/0.06/0.00/0.17  \\

\bottomrule[1pt]\bottomrule[1pt]
\end{tabular}
\end{table}

\renewcommand\tabcolsep{0.45cm}
\renewcommand{\arraystretch}{1.50}
\begin{table}[!h]
\caption{Numerical solutions for the bound states with the $S-D$ wave mixing in the $B_s\bar{B}_{s1}$ system. Conventions are the same as those in Table~\ref{sr1}.}\label{sr9}
\begin{tabular}{c|cccl}\toprule[1.0pt]\toprule[1.0pt]
$J^{PC}$&$\Lambda$      &$E$&$r_{\rm RMS}$&$P({}^3\mathbb{S}_{1}/{}^3\mathbb{D}_{1})$  \\\midrule[1.0pt]
\multirow{3}{*}{$1^{--}$}         &1.87&$0.15$&3.73&\textbf{100}/0.00  \\
                                    &2.01&$1.82$&1.60&\textbf{100}/0.00  \\
                                    &2.15&$4.97$&1.04&\textbf{100}/0.00\\
                                    \midrule[1.0pt]
\multirow{3}{*}{$1^{-+}$}         &1.80&$0.14$&3.75&\textbf{100}/0.00  \\
                                    &1.93&$1.87$&1.59&\textbf{100}/0.00 \\
                                    &2.06&$5.16$&1.03&\textbf{100}/0.00 \\
                            
\bottomrule[1pt]\bottomrule[1pt]
\end{tabular}
\end{table}
\renewcommand\tabcolsep{0.40cm}
\renewcommand{\arraystretch}{1.50}
\begin{table}[!htbp]
\caption{Numerical solutions for the bound states with the $S-D$ wave mixing in the $B^*_s\bar{B}_{s1}$  system. Conventions are the same as those in Table~\ref{sr1}.}\label{sr10}
\begin{tabular}{c|cccl}\toprule[1.0pt]\toprule[1.0pt]
$J^{PC}$&$\Lambda$      &$E$&$r_{\rm RMS}$&$P({}^1\mathbb{S}_{0}/{}^5\mathbb{D}_{0})$  \\
\multirow{3}{*}{$0^{--}$}         &1.29&$0.37$&2.88&\textbf{99.86}/0.14  \\
                                    &1.31&$1.89$&1.67&\textbf{99.79}/0.21  \\
                                    &1.33&$4.50$&0.99&\textbf{99.77}/0.23 \\                      
                                    \midrule[1.0pt]
\multirow{3}{*}{$0^{-+}$}         &1.29&$0.49$&2.61&\textbf{99.85}/0.15 \\
                                    &1.31&$2.14$&1.38&\textbf{99.79}/0.21 \\
                                    &1.33&$4.88$&0.96&\textbf{99.77}/0.23 \\
                                    \midrule[1.0pt]
$J^{PC}$&$\Lambda$ &$E$&$r_{\rm RMS}$&$P({}^3\mathbb{S}_{1}/{}^3\mathbb{D}_{1}/{}^5\mathbb{D}_{1})$ \\                                  
\multirow{3}{*}{$1^{--}$}         &1.43&$0.21$&3.41&\textbf{99.86}/0.14/0.00  \\
                                    &1.46&$1.51$&1.65&\textbf{99.76}/0.24/0.00  \\
                                    &1.49&$3.95$&1.08&\textbf{99.72}/0.28/0.00 \\                      
                                    \midrule[1.0pt]
\multirow{3}{*}{$1^{-+}$}         &1.43&$0.29$&3.15&\textbf{99.85}/0.15/0.00  \\
                                    &1.46&$1.70$&1.56&\textbf{99.75}/0.25/0.00 \\
                                    &1.49&$4.26$&1.04&\textbf{99.72}/0.28/0.00 \\  
                                    \midrule[1.0pt]   
$J^{PC}$&$\Lambda$ &$E$&$r_{\rm{RMS}}$&$P({}^5\mathbb{S}_{2}/{}^1\mathbb{D}_{2}/{}^3\mathbb{D}_{2}/{}^5\mathbb{D}_{2})$ \\
\multirow{3}{*}{$2^{--}$}           &2.30&$0.20$&3.63&\textbf{99.85}/0.12/0.00 /0.13\\
                                    &2.65&$1.87$&1.71&\textbf{99.78}/0.02/0.00/0.20 \\
                                    &3.00&$5.10$&1.16&\textbf{99.77}/0.06/0.00/0.17 \\  
                                    \midrule[1.0pt]  
\multirow{1}{*}{$2^{-+}$}
                                    &2.30&$0.14$&3.89&\textbf{99.87}/0.02/0.00 /0.20\\
                                    &2.65&$1.66$&1.80&\textbf{99.78}/0.02/0.00/0.20 \\
                                    &3.00&$4.70$&1.20&\textbf{99.77}/0.06/0.00/0.17 \\
\bottomrule[1pt]\bottomrule[1pt]
\end{tabular}
\end{table}
\renewcommand\tabcolsep{0.35cm}
\renewcommand{\arraystretch}{1.50}
\begin{table}[!hbtp]
\caption{Numerical solutions for the bound states with the $S-D$ wave mixing in the  $B_s^{*}\bar B^*_{s2}$  system. Conventions are the same as those in Table~\ref{sr1}.}\label{sr11}
\begin{tabular}{c|cccl}\toprule[1.0pt]\toprule[1.0pt]
$J^{PC}$&$\Lambda$&$E$&$r_{\rm RMS}$&$P({}^3\mathbb{S}_{1}/{}^3\mathbb{D}_{1}/{}^5\mathbb{D}_{1}/{}^7\mathbb{D}_{1})$ \\
\multirow{3}{*}{$1^{--}$}         &1.31&$0.22$&3.33&\textbf{99.94}/0.02/0.00/0.04 \\
                                    &1.33&$1.48$&1.63&\textbf{99.90}/0.03/0.00/0.07  \\
                                    &1.35&$3.76$&1.07&\textbf{99.89}/0.04/0.00/0.07  \\                     
                                    \midrule[1.0pt]             
\multirow{3}{*}{$1^{-+}$}         &1.31&$0.28$&3.15&\textbf{99.94}/0.02/0.00/0.04  \\
                                    &1.33&$1.61$&1.57&\textbf{99.90}/0.03/0.00/0.07  \\
                                    &1.35&$3.95$&1.05&\textbf{99.89}/0.04/0.00/0.07  \\
                                    \midrule[1.0pt]                
$J^{PC}$&$\Lambda$&$E$&$r_{\rm RMS}$&$P({}^5\mathbb{S}_{2}/{}^3\mathbb{D}_{2}/{}^5\mathbb{D}_{2}/{}^7\mathbb{D}_{2})$ \\   
\multirow{3}{*}{$2^{--}$}         &1.54&$0.19$&3.49&\textbf{99.92}/0.00/0.08/0.00  \\
                                    &1.59&$1.92$&1.50&\textbf{99.86}/0.00/0.14/0.00\\
                                    &1.63&$4.56$&1.03&\textbf{99.84}/0.00/0.16/0.00  \\
                                    \midrule[1.0pt]                          
\multirow{3}{*}{$2^{-+}$}         &1.52&$0.12$&3.80&\textbf{99.93}/0.00/0.07/0.00  \\
                                    &1.56&$1.25$&1.82&\textbf{99.87}/0.00/0.13/0.00  \\
                                    &1.61&$4.24$&1.06&\textbf{99.84}/0.00/0.16/0.00  \\
                                    \midrule[1.0pt] 

$J^{PC}$&$\Lambda$&$E$&$r_{\rm RMS}$&$P({}^7\mathbb{S}_{3}/{}^3\mathbb{D}_{3}/{}^5\mathbb{D}_{3}/{}^7\mathbb{D}_{3})$ \\   
\multirow{3}{*}{$3^{--}$}        
                                    &2.35&$0.19$&3.68&\textbf{99.86}/0.01/0.00/0.13  \\
                                    &2.70&$1.89$&1.72&\textbf{99.78}/0.01/0.00/0.21  \\
                                    &3.00&$4.60$&1.22&\textbf{99.80}/0.02/0.00/0.18  \\
                                    \midrule[1.0pt] 
\multirow{1}{*}{$3^{-+}$}          &2.20&$0.17$&3.75&\textbf{99.85}/0.01/0.00/0.14  \\
                                    &2.60&$2.47$&1.55&\textbf{99.75}/0.01/0.00/0.24  \\
                                    &3.00&$7.04$&1.05&\textbf{99.81}/0.02/0.00/0.17  \\             
\bottomrule[1pt]\bottomrule[1pt]
\end{tabular}
\end{table}

\renewcommand\tabcolsep{0.40cm}
\renewcommand{\arraystretch}{1.50}
\begin{table}[!hbt]
\caption{Numerical solutions for the bound states with the $S-D$ wave mixing in the $H_s T_s$ systems. Conventions are the same as those in Table~\ref{sr1}.}\label{sr12}
\begin{tabular}{c|cccl}\toprule[1pt]\toprule[1pt]
\multicolumn{5}{c}{$B_s^{*} B_{s1}$}\\
\cline{1-5}
$J^{P}$&$\Lambda$ &$E$&$r_{\rm RMS}$&$P({}^1\mathbb{S}_{0}/{}^5\mathbb{D}_{0})$ \\
\multirow{3}{*}{$0^{-}$} 
&1.62&$0.12$&3.86&\textbf{98.65}/1.35\\
&1.66&$1.45$&1.78&\textbf{96.19}/3.82\\
&1.70&$4.46$&1.10&\textbf{93.81}/6.19\\
\cline{1-5}
\multicolumn{5}{c}{$B_s^{*} B_{s1}$}\\
\cline{1-5}
$J^{P}$&$\Lambda$ &$E$&$r_{\rm RMS}$&$P({}^3\mathbb{S}_{1}/{}^3\mathbb{D}_{1}/{}^5\mathbb{D}_{1})$ \\
\multirow{3}{*}{$1^{-}$} 
&1.63&$0.19$&3.56&\textbf{98.86}/1.14/0.00\\
&1.68&$1.62$&1.70&\textbf{97.12}/2.88/0.00\\
&1.74&$5.34$&1.03&\textbf{95.16}/4.84/0.00\\
\cline{1-5}
\multicolumn{5}{c}{$B_s^{*} B_{s1}$}\\
\cline{1-5}
$J^{P}$&$\Lambda$ &$E$&$r_{\rm RMS}$&$P({}^5\mathbb{S}_{2}/{}^1\mathbb{D}_{2}/{}^3\mathbb{D}_{2}/{}^5\mathbb{D}_{2})$ \\
\multirow{3}{*}{$1^{-}$} 
&1.53&$0.14$&3.73&\textbf{98.97}/0.17/0.00/0.85\\
&1.62&$2.30$&1.47&\textbf{96.62}/0.60/0.00/2.78\\
&1.70&$6.13$&1.00&\textbf{94.63}/0.98/0.00/4.39\\
\cline{1-5}
\multicolumn{5}{c}{$B_s^{*} B_{s2}^{*}$} \\
\cline{1-5}
$J^{P}$&$\Lambda$ &$E$&$r_{\rm RMS}$&$P({}^3\mathbb{S}_{1}/{}^3\mathbb{D}_{1}/{}^5\mathbb{D}_{1}/{}^7\mathbb{D}_{1})$ \\
\multirow{3}{*}{$1^{-}$} 
&1.70&$0.20$&3.52&\textbf{98.28}/0.12/0.00/0.60\\
&1.74&$1.56$&1.7&\textbf{98.23}/0.28/0.00/4.49\\
&1.78&$4.45$&1.07&\textbf{97.25}/0.40/0.00/2.34\\
\cline{1-5}
\multicolumn{5}{c}{$B_s^{*} B_{s2}^{*}$} \\
\cline{1-5}
$J^{P}$&$\Lambda$ &$E$&$r_{\rm RMS}$&$P({}^5\mathbb{S}_{2}/{}^3\mathbb{D}_{2}/{}^5\mathbb{D}_{2}/{}^7\mathbb{D}_{2})$ \\
\multirow{3}{*}{$2^{-}$} 
&1.71&$0.13$&3.79&\textbf{99.51}/0.00/0.49/0.00\\
&1.78&$1.62$&1.68&\textbf{99.61}/0.00/1.39/0.00\\
&1.86&$5.41$&1.01&\textbf{97.65}/0.00/2.35/0.00\\
\cline{1-5}
\multicolumn{5}{c}{$B_s^{*} B_{s2}^{*}$} \\
\cline{1-5}
$J^{P}$&$\Lambda$ &$E$&$r_{\rm RMS}$&$P({}^7\mathbb{S}_{3}/{}^3\mathbb{D}_{3}/{}^5\mathbb{D}_{3}/{}^7\mathbb{D}_{3})$ \\
\multirow{3}{*}{$3^{-}$} 
&1.56&$3.67$&0.16&\textbf{99.02}/0.12/0.00/0.86\\
&1.65&$1.99$&1.56&\textbf{97.06}/0.37/0.00/2.57\\
&1.75&$6.12$&1.00&\textbf{95.00}/0.67/0.00/4.33\\
\bottomrule[1pt]\bottomrule[1pt]
\end{tabular}
\end{table}

In the $B^*B_{1}$ system, for the $I(J^{P})=0(0^{-})$ state, when $\Lambda$ is set to be $1320~\rm{MeV}$, $1370~\rm{MeV}$, and $1420~\rm{MeV}$, the corresponding binding energies $E$ are $0.16~\rm{MeV}$, $2.39~\rm{MeV}$, and $7.05~\rm{MeV}$, and the $r_{\rm RMS}$ are $3.64~\rm{fm}$, $1.49~\rm{fm}$, and $1.05~\rm{fm}$, respectively. The dominant probabilities are $97.55\%$, $96.20\%$, and $96.14\%$, are concentrated in the ${}^1\mathbb{S}_{0}$ channel, while the ${}^1\mathbb{D}_{0}$ channel contributes smaller probabilities of $2.45\%$, $3.80\%$, and $3.86\%$, respectively. For the $I(J^{P})=1(0^{-})$ state, no bound state is observed. Additionally, for the $I(J^{P})=0(1^{-})$ configuration can also be regarded  as a molecular state, when $\Lambda$ varies from $840~\rm{MeV}$ to $960~\rm{MeV}$, the corresponding binding energy $E$ is about  $0.31\sim 6.52~\rm{MeV}$. Moreover, bound state solutions exist for both $I(J^{P})=0(2^-)$ and $1(2^-)$ configurations. However, the bound state in the  $1(2^-)$ state appears only for large values of the cutoff parameter, and thus we do not consider it a viable candidate for a molecular bound state. 

In the $B^*B^*_{2}$ system, for the $I(J^{P})=0(1^{-})$ state, when the cutoff parameter $\Lambda$ is set to be $800\sim810~\rm{MeV}$, the corresponding binding energies $E$ are $5.29~\rm{MeV}$ and $6.06~\rm{MeV}$, and the root-mean-square radii $r_{\rm RMS}$ are $1.08~\rm{fm}$ and $1.02~\rm{fm}$, respectively. The dominant probabilities are concentrated in the ${}^3\mathbb{S}_{1}$ channel, with values of $99.29\%$ and $99.30\%$, respectively. For the $I(J^{P})=0(2^{-})$ state, we vary the cutoff in a range of $930\sim1060~\rm MeV$, the root-mean-square radii $r_{\rm RMS}$ are $3.89\sim1.01~\rm{fm}$, which can be regarded as a good molecular state. Furthermore, for the $I(J^{P})=0(3^{-})$ state, we obtain a loosely bound state when the cutoff varies in the range of $1180\sim1840~\rm MeV$. Also, we find a bound state for the $I(J^{P})=1(3^{-})$ $B^*B^*_{2}$ channel that has higher isospin and spin.

For all the bound states in the $HT$ systems, the dominant components belong to the $S$-wave channel, while the $D$-wave components are small. The dominance of the $S$-wave channel indicates that these states possess significant spatial overlap, facilitating the formation of molecular states. Moreover, it can be seen that with the rise of cutoff parameter $\Lambda$, the binding energy deepens progressively, and the root-mean-square radius decreases, which is a typical feature of present results. 

Moreover, for the isospin $I=0$ channels, including $BB_{1}$, $BB^*_{2}$, $B^*B_{1}$, and $B^*B^*_{2}$, bound states appear around $\Lambda=1000~\rm{MeV}$, indicating that the $I=0$ $HT$ system are promising candidates for hidden-bottom tetraquark molecular states. Also, the $1(2^-)$ state in the $B^*B_{1}$ channel and the $1(3^-)$ state in the $B^*B^*_{2}$ channel for isospin $I=1$ may exist as bound states. The absence of a bound state in some $I=1$ channels indicates that weakly attractive or even repulsive potentials appear in these higher isospin configurations.


\subsection{The $H_s\bar{T}_s$ and $H_sT_s$ systems}

For the $H_s\bar{T}_s$ and $H_sT_s$ systems, when we vary the cutoff in a range  $800\sim1200~\rm MeV$, no bound state poles are observed, which suggests that interactions between two bottom-strange mesons of $H_s$ and $T_s$ types may be insufficient to form a bound molecular state. For instance, the $B_s\bar{B}_{s1}$ system shows no bound state solution within the relevant $\Lambda$ range. Given the heavy quark flavor symmetry, it can be conjectured that its charmed analogue $D_s\bar{D}_{s1}$ system cannot form a bound state. Recently, the BESIII Collaboration investigated the process $e^+e^- \to \gamma D^+_s D_{s1}^-(2536) + \text{c.c.}$~\cite{BESIII:2025vea} in a search for an exotic molecular state with quantum numbers $J^{PC}=1^{-+}$, but no significant signal was observed. Our current findings align with this experimental result. 
 
Theoretically,we can extend the cutoff range to $1200\sim3000~\rm MeV$ to investigate the behavior of these channels, which may offer some useful information for $H_sT_s$ and $H_s\bar{T}_s$ systems. In the $I(J^{PC})=1^{-\pm} B_s\bar{B}_{s1}$ channels, we obtain two bound state solutions at the cutoff around $1800~\rm MeV$. For the $B^*_s\bar{B}_{s1}$ channel, we obtain two exotic structures for $0^{-\pm}$ states at cutoffs in the range  $1290\sim1330~\rm MeV$, where the $r_{RMS}$ is $3\sim1~\rm fm$. Also, for $1^{-\pm}$ states, two bound states could exist when we increase the cutoff to  $1430~\rm MeV$. For the $B^*_s\bar{B}^*_{s2}$ channel, multiple bound states were also observed for the quantum numbers $J^{PC}=1^{-\pm}$ at larger cutoff values. However, all these bound solutions need significantly large $\Lambda$ that is far away from $1000~\rm{MeV}$ scale, and thus may not correspond to genuine molecular states.

\section{Summary}\label{sec4}
\begin{table*}[]
\caption{{Numerical solutions for the S-D wave mixing bound states in the $H\bar{T}$ and $HT$ systems, obtained with the same values of the cutoff parameter $\Lambda$. The units for $\Lambda$, binding energy $E$, and RMS radius $r_{\mathrm{RMS}}$ are $\mathrm{GeV}$, $\mathrm{MeV}$, and $\mathrm{fm}$, respectively. The ``$\times$" denotes the absence of a bound state for $\Lambda$ scanned from 0.9 to 1.1~$\mathrm{GeV}$. The symbol $\ddots$ indicates the absence of a state with the corresponding quantum numbers. }}\label{srt1}
{
\begin{tabular}{cc|cc|cc|cc|cc}
\toprule[1pt]\toprule[1pt]
&&\multicolumn{2}{c}{$B{\bar{B}_{1}}$}&\multicolumn{2}{c}{$B \bar{B}^*_{2}$}&\multicolumn{2}{c}{$B^*{\bar{B}_{1}}$}&\multicolumn{2}{c}{$B^*{\bar B^*_{2}}$}\\
\hline
$J^{P}$&$\Lambda$ &$E$&$r_{\rm RMS}$&$E$&$r_{\rm RMS}$&$E$&$r_{\rm RMS}$&$E$&$r_{\rm RMS}$\\
\hline
\multirow{3}{*}{$0^{--}$} 
& 0.9 & \multicolumn{2}{c|}{\multirow{3}{*}{$\ddots$}} & \multicolumn{2}{c|}{\multirow{3}{*}{$\ddots$}} & $\times$ &$\times$& \multicolumn{2}{c}{\multirow{3}{*}{$\ddots$}} \\
& 1.0 & \multicolumn{2}{c|}{} & \multicolumn{2}{c|}{} & $\times$ &$\times$& \multicolumn{2}{c}{} \\
& 1.1 & \multicolumn{2}{c|}{} & \multicolumn{2}{c|}{} & $\times$ &$\times$& \multicolumn{2}{c}{} \\
\hline
\multirow{3}{*}{$0^{-+}$} 
& 0.9 & \multicolumn{2}{c|}{\multirow{3}{*}{$\ddots$}} & \multicolumn{2}{c|}{\multirow{3}{*}{$\ddots$}} & $\times$ &$\times$& \multicolumn{2}{c}{\multirow{3}{*}{$\ddots$}} \\
& 1.0 & \multicolumn{2}{c|}{} & \multicolumn{2}{c|}{} & $\times$ &$\times$& \multicolumn{2}{c}{} \\
& 1.1 & \multicolumn{2}{c|}{} & \multicolumn{2}{c|}{} & $\times$ &$\times$& \multicolumn{2}{c}{} \\
\hline
\multirow{3}{*}{$1^{--}$} 
& 0.9 & $\times$&$\times$ & \multicolumn{2}{c|}{\multirow{3}{*}{$\ddots$}} & $-5.63$  &$1.10$  & $\times$& $\times$\\
& 1.0 & $\times$&$\times$ & \multicolumn{2}{c|}{} & $\times$ &$\times$& $\times$& $\times$ \\
& 1.1 &$-2.01$    & 1.64    & \multicolumn{2}{c|}{} & $\times$ &$\times$& $\times$& $\times$\\
\hline
\multirow{3}{*}{$1^{-+}$} 
& 0.9 & $\times$&$\times$ & \multicolumn{2}{c|}{\multirow{3}{*}{$\ddots$}} & $-5.17$  &$1.132$ & $\times$ & $\times$ \\
& 1.0 & $\times$&$\times$ & \multicolumn{2}{c|}{} & $\times$ &$\times$& $\times$ & $\times$\\
& 1.1 &$-1.51$    &    1.89 & \multicolumn{2}{c|}{} & $\times$ &$\times$& $\times$ & $\times$\\
\hline
\multirow{3}{*}{$2^{--}$} 
& 0.9 & \multicolumn{2}{c|}{\multirow{3}{*}{$\ddots$}} & $\times$ &$\times$  & $\times$ &$\times$ & $-0.36$ & $3.11$\\
& 1.0 & \multicolumn{2}{c|}{} & $\times$ & $\times$ & $\times$ &$\times$ & $-9.73$ & $0.89$ \\
& 1.1 & \multicolumn{2}{c|}{} & $-1.68$ & 1.79 &$-0.76$ &$2.77$& $\times$& $\times$\\
\hline
\multirow{3}{*}{$2^{-+}$} 
& 0.9 & \multicolumn{2}{c|}{\multirow{3}{*}{$\ddots$}} & $\times$ & $\times$ & $\times$ &$\times$ & $\times$ & $\times$ \\
& 1.0 & \multicolumn{2}{c|}{} & $\times$ & $\times$ & $\times$ &$\times$ & $-7.30$ & $0.98$\\
& 1.1 & \multicolumn{2}{c|}{} & $-1.68$ & 1.79 & $-0.52$ &3.14& $\times$ & $\times$\\
\hline
\multirow{3}{*}{$3^{--}$} 
& 0.9 & \multicolumn{2}{c|}{\multirow{3}{*}{$\ddots$}} & \multicolumn{2}{c|}{\multirow{3}{*}{$\ddots$}} & \multicolumn{2}{c|}{\multirow{3}{*}{$\ddots$}} & $\times$ &$\times$  \\
& 1.0 & \multicolumn{2}{c|}{} & \multicolumn{2}{c|}{} & \multicolumn{2}{c|}{} & $\times$ & $\times$ \\
& 1.1 & \multicolumn{2}{c|}{} & \multicolumn{2}{c|}{} & \multicolumn{2}{c|}{} & $-0.56$ &$3.09$  \\
\hline
\multirow{3}{*}{$3^{-+}$} 
& 0.9 & \multicolumn{2}{c|}{\multirow{3}{*}{$\ddots$}} & \multicolumn{2}{c|}{\multirow{3}{*}{$\ddots$}} & \multicolumn{2}{c|}{\multirow{3}{*}{$\ddots$}} & $\times$ &$\times$  \\
& 1.0 & \multicolumn{2}{c|}{} & \multicolumn{2}{c|}{} & \multicolumn{2}{c|}{} & $\times$ & $\times$ \\
& 1.1 & \multicolumn{2}{c|}{} & \multicolumn{2}{c|}{} & \multicolumn{2}{c|}{} & $0.44$ &$3.34$  \\
\hline
&&\multicolumn{2}{c}{$B{{B}_{1}}$}&\multicolumn{2}{c}{$B {B}^*_{2}$}&\multicolumn{2}{c}{$B^*{{B}_{1}}$}&\multicolumn{2}{c}{$B^*{ B^*_{2}}$}\\
\hline
$J^{P}$&$\Lambda$ &$E$&$r_{\rm RMS}$&$E$&$r_{\rm RMS}$&$E$&$r_{\rm RMS}$&$E$&$r_{\rm RMS}$\\
\hline
\multirow{3}{*}{$0^{-}$} 
& 0.9 & \multicolumn{2}{c|}{\multirow{3}{*}{$\ddots$}} & \multicolumn{2}{c|}{\multirow{3}{*}{$\ddots$}} & $\times$ &$\times$& \multicolumn{2}{c}{\multirow{3}{*}{$\ddots$}} \\
& 1.0 & \multicolumn{2}{c|}{} & \multicolumn{2}{c|}{} & $\times$ &$\times$& \multicolumn{2}{c}{} \\
& 1.1 & \multicolumn{2}{c|}{} & \multicolumn{2}{c|}{} & $\times$ &$\times$& \multicolumn{2}{c}{} \\
\hline
\multirow{3}{*}{$1^{-}$} 
& 0.9 & $\times$&$\times$ & \multicolumn{2}{c|}{\multirow{3}{*}{$\ddots$}} & $-2.19$  &$1.58$  & $\times$& $\times$\\
& 1.0 & $\times$&$\times$ & \multicolumn{2}{c|}{} & $-11.13$ &$0.83$& $\times$& $\times$ \\
& 1.1 & $\times$&$\times$ & \multicolumn{2}{c|}{} & $\times$ &$\times$& $\times$& $\times$\\
\hline
\multirow{3}{*}{$2^{-}$} 
& 0.9 & \multicolumn{2}{c|}{\multirow{3}{*}{$\ddots$}} & $\times$ & $\times$ & $-0.44$ &$3.27$ & $\times$ & $\times$\\
& 1.0 & \multicolumn{2}{c|}{} & $\times$ & $\times$ & $-4.72$ &$1.41$ & $-2.37$ & $1.52$ \\
& 1.1 & \multicolumn{2}{c|}{} & $\times$ & $\times$ &$-12.11$ &$1.02$& $-11.145$& $0.83$\\
\hline
\multirow{3}{*}{$3^{-}$} 
& 0.9 & \multicolumn{2}{c|}{\multirow{3}{*}{$\ddots$}} & \multicolumn{2}{c|}{\multirow{3}{*}{$\ddots$}} & \multicolumn{2}{c|}{\multirow{3}{*}{$\ddots$}} & $\times$ &$\times$  \\
& 1.0 & \multicolumn{2}{c|}{} & \multicolumn{2}{c|}{} & \multicolumn{2}{c|}{} & $\times$ & $\times$ \\
& 1.1 & \multicolumn{2}{c|}{} & \multicolumn{2}{c|}{} & \multicolumn{2}{c|}{} & $\times$ &$\times$  \\
\bottomrule[1pt]\bottomrule[1pt]
\end{tabular}
}
\end{table*}
In this work, we perform a systematic study on the doubly bottom  $H_{(s)}\bar{T}_{(s)}$  and $H_{(s)}T_{(s)}$ systems within the one-boson-exchange (OBE) model.
When the cutoff parameter $\Lambda$ is around $1000~\MeV$, we obtain some loosely bound states in the $H\bar{T}$ and $HT$ systems . 
That is, the $I(J^{PC})=0(1^{-\pm})$ $B\bar{B}_{1}$, $I(J^{PC})=0(2^{-\pm})$ $B\bar{B}_{2}^{*}$, $I(J^{PC})=0(1^{-\pm})$ $B^*\bar{B}_{1}$, and $I(J^{PC})=0(2^{-\pm})$ $B^*\bar{B}_{2}^{*}$ configurations in the $H\bar{T}$ system are the most promising hidden bottom molecular tetraquark candidates,  the  $B^*B_1$ channels with $I(J^{PC})=0(1^{-}), 0(2^{-})$ and the $B^*B_2^*$ channel with $I(J^{PC})=0(2^{-})$ are also likely candidates for forming molecular tetraquarks. These predictions together with possible decay modes are collected in Table~\ref{srt1}. 
Finally, we cannot find any molecular candidate in the bottom-strange sectors. 
Particularly, with a reasonable cutoff parameter, the $B_s \bar B_{s1}$ system cannot form a bound state within our framework, and its charmed analog, the $D_s \bar D_{s1}$ state was also not observed in a recent measurement by the BESIII Collaboration. 
\begin{table*}[htb]
 \renewcommand\tabcolsep{0.2cm}
 	\renewcommand\arraystretch{1.5}
 	\caption{ The summary of our predictions for possible molecular states of $H\bar{T}(T)$ systems and their possible decay modes with cutoff $\Lambda$ in a range of $900\sim1100$ $\rm~MeV$.}\label{srt2}
 			\begin{tabular}{cccccccc}\toprule[1.0pt]\toprule[1.0pt]
 			&$I(J^{PC})$& System&Mass$(\rm MeV)$&$r_{RMS}(\rm fm)$
 			&Possible decay mode
 			\\\hline
 			&\multirow{2}{*}{{$0(1^{--})$}}
    &{$ B\bar{B}_1+h.c.$}&{$10998.70\sim11005.00$}&{$1.03\sim3.12$} &{$\eta_b(nS) \eta^{(\prime)}/\eta_b(nS) \omega/\Upsilon(nS) \omega/\Upsilon(nS) \eta^{(\prime)}/B\bar{B}/B\bar{B}^*+h.c./B^*\bar{B}^*$}\\
    
    &&{$ B^* \bar{B}_1+h.c.$}&{$11043.60\sim11050.40$}&{$1.01\sim3.62$} &{$\eta_b(nS) \eta^{(\prime)}/\eta_b(nS) \omega/\Upsilon(nS) \omega/\Upsilon(nS) \eta^{(\prime)}/B\bar{B}/B\bar{B}^*+h.c./B^*\bar{B}^*$}
    \\
    &\multirow{2}{*}{{$0(1^{-+})$}}
    &{$ B\bar{B}_1+h.c.$}&{$10998.40\sim11005.20$}&{$1.04\sim3.55$} &{$\eta_b(nS) \eta^{(\prime)}/\eta_b(nS) \omega/\Upsilon(nS) \omega/\Upsilon(nS) \eta^{(\prime)}/B\bar{B}/B\bar{B}^*+h.c./B^*\bar{B}^*$}\\
    &&{$ B^* \bar{B}_1+h.c.$}&{$11044.20\sim11050.50$}&{$1.04\sim3.87$} &{$\eta_b(nS) \eta^{(\prime)}/\eta_b(nS) \omega/\Upsilon(nS) \omega/\Upsilon(nS) \eta^{(\prime)}/B\bar{B}/B\bar{B}^*+h.c./B^*\bar{B}^*$}
    \\
    
  &\multirow{2}{*}{{$0(2^{--})$}}
  &{$B\bar{B}^*_2+h.c.$}&{$11010.00\sim11016.40$}&{$1.04\sim3.41$} &{$\eta_b(nS) \omega/\Upsilon(nS) \omega/ \Upsilon(nS) \eta^{(\prime)}/B\bar{B}^*+h.c./B^*\bar{B}^*$}\\
 &&{$B^*\bar{B}^*_2+h.c.$}&{$11055.20\sim11061.80$}&{$1.02\sim3.86$} &{$\eta_b(nS) \omega/\Upsilon(nS) \omega/ \Upsilon(nS) \eta^{(\prime)}/B\bar{B}^*+h.c./B^*\bar{B}^*$}\\
 &\multirow{2}{*}{{$0(2^{-+})$}}
  &{$B\bar{B}^*_2+h.c.$}&{$11010.00\sim11016.40$}&{$1.04\sim3.41$} &{$\eta_b(nS) \omega/\Upsilon(nS) \omega/ \Upsilon(nS) \eta^{(\prime)}/B\bar{B}^*+h.c./B^*\bar{B}^*$}\\
 &&{$B\bar{B}^*_2+h.c.$}&{$11055.90\sim11061.70$}&{$1.07\sim3.48$} &{$\eta_b(nS) \omega/\Upsilon(nS) \omega/ \Upsilon(nS) \eta^{(\prime)}/B\bar{B}^*+h.c./B^*\bar{B}^*$}\\
  
  &\multirow{1}{*}{{$0(1^{-})$}}	
  &{$B^*{B}_1+h.c.$}&{$11044.10\sim11050.30$}&{$1.03\sim3.26$} &{$B{B}^*+h.c./B^*{B}^*$}\\
    &\multirow{1}{*}{{$0(2^{-})$}}	
  &{$B^*{B}_1+h.c.$}&{$11038.50\sim11050.20$}&{$1.02\sim3.27$} &{$B{B}^*+h.c./B^*{B}^*$}\\
 
   &\multirow{1}{*}{{$0(2^{-})$}}
  &{$B^*{B}^*_2+h.c.$}&{$11055.10\sim11061.80$}&{$1.01\sim3.89$} &{$B{B}^*+h.c./B^*{B}^*$ }\\
   
   \bottomrule[1pt]\bottomrule[1pt]
 		\end{tabular}
 \end{table*}

{In this work, we perform a systematic study on the doubly bottom  $H_{(s)}\bar{T}_{(s)}$  and $H_{(s)}T_{(s)}$ systems within the one-boson-exchange (OBE) model.
When the cutoff parameter $\Lambda$ is around $1000~\MeV$, we obtain some loosely bound states in the $H\bar{T}$ and $HT$ systems . 
That is, the $I(J^{PC})=0(1^{-\pm})$ $B\bar{B}_{1}$, $I(J^{PC})=0(2^{-\pm})$ $B\bar{B}_{2}^{*}$, $I(J^{PC})=0(1^{-\pm})$ $B^*\bar{B}_{1}$, and $I(J^{PC})=0(2^{-\pm})$ $B^*\bar{B}_{2}^{*}$ configurations in the $H\bar{T}$ system are the most promising hidden bottom molecular tetraquark candidates,  the  $B^*B_1$ channels with $I(J^{PC})=0(1^{-}), 0(2^{-})$ and the $B^*B_2^*$ channel with $I(J^{PC})=0(2^{-})$ are also likely candidates for forming molecular tetraquarks. To clearly elucidate the properties of the molecular state candidates and to facilitate a comparison among them, we calculated the bound states using a uniform cutoff parameter. The results are detailed in Table~\ref{srt1}. Furthermore, the corresponding theoretical predictions and possible decay modes are summarized in Table~\ref{srt2}.}
Finally, we cannot find any molecular candidate in the bottom-strange sectors. 
Particularly, with a reasonable cutoff parameter, the $B_s \bar B_{s1}$ system cannot form a bound state within our framework, and its charmed analog, the $D_s \bar D_{s1}$ state was also not observed in a recent measurement by the BESIII Collaboration.

According to our calculations, we find that the $I=0$ configurations can form  bound states more easily than the $I=1$ configurations, and the bound states tend to emerge in the non-strange $H\bar{T}$  and $HT$ systems. 
Also, the $D$-wave components play an essential role in some bound states as well as the deuteron. Our aim is to establish the likely existence of bound states in specific channels and provide reasonable estimates of binding energies, thereby yielding approximate masses for molecular states to guide experimental searches. The accelerated discovery of new particles can, in turn, facilitate the progressive refinement of phenomenological models. We encourage more experimental and theoretical researchers to participate in the study of these doubly bottom molecular states.

\section*{ACKNOWLEDGMENTS}

We would like to thank Fu-Lai Wang for his valuable discussion, which provided a significant inspiration for our work. This research was supported by the start-up research fund of Henan University. Q.-F. L\"u is supported by Hunan Provincial Natural Science Foundation of China under Grant No. 2023JJ40421, the Scientific Research Foundation of Hunan Provincial Education Department under Grant No. 24B0063, and the Youth Talent Support Program of Hunan Normal University under Grant No. 2024QNTJ14. 
     
\appendix
\section{The relevant parameters and potentials}\label{app:1}

\renewcommand\tabcolsep{0.13cm}
\renewcommand{\arraystretch}{1.50}
\begin{table}[!ht]
\caption{A summary of coupling constants and masses for mesons. The coupling constants $h^{\prime} = (h_1+h_2)/\Lambda_{\chi}$, $\lambda$, $\lambda^{\prime\prime}$, and $\mu_1$ are listed in unit of $\rm{GeV}^{-1}$, $f_{\pi}$ in $\rm{GeV}$, and masses in $\rm{MeV}$.}\label{parameters}
\centering
\begin{tabular}{llll}
\toprule[1.0pt]
$g_{\sigma}=-0.76$     & $g^{\prime\prime}_{\sigma}=0.76$  & $|h^{\prime}_{\sigma}|=0.35$ & $g=0.59$ \\
$k=-0.59$              & $|h^{\prime}|=0.55$              & $f_{\pi}=0.132$              & $\beta=-0.90$ \\
$\beta^{\prime\prime}=0.90$ & $\lambda=-0.56$          & $\lambda^{\prime\prime}=0.56$ & $\zeta_1=0.20$ \\
$\zeta=0.73$           & $\mu=0.36$                 & $g_V=5.83$                   & $m_{\sigma}=600.00$ \\     
$m_{\pi}=137.27$       & $m_{\eta}=547.86$          & $m_{\rho}=775.26$            & $m_{\omega}=782.66$ \\
$m_B=5729.50$          & $m_{B^*}=5324.70$          & $m_{B_{1}}=5725.90$          & $m_{B_{2}^{*}}=5737.20$ \\  
$m_{B_{s}}=5366.87$    & $m_{B_{s}^*}=5415.55$      & $m_{B_{s1}}=5828.70$         & $m_{B_{s2}^{*}}=5839.86$\\
\bottomrule[1.0pt]
\end{tabular}
\end{table}

\renewcommand\tabcolsep{0.13cm}
\renewcommand{\arraystretch}{1.50}
\begin{table}[!ht]
	\renewcommand\arraystretch{1.6}
	\caption{\label{zb}  The effective  potentials for the $H_{({s})}\bar{T}_{({s})}$ systems.}
	\begin{ruledtabular}
		\begin{tabular}{ccccccccc}
			&System
			&Effective potential\\\hline
      &\multirow{2}{*}{$V_{B\bar{B}_{1}\to B\bar{B}_{1}} $}
      &$_\sigma g_\sigma''\mathcal{O}_1Y_\sigma+\frac{1}{2}\beta\beta''g_V^2\mathcal{O}_1\mathcal{G}(I)Y_V$\\
      &&$+c \times \left[\frac{2h_\sigma^{\prime2}}{9f_\pi^2}\left(\mathcal{O}_2\mathcal{Z}+\mathcal{O}_3\mathcal{T}\right)Y_{\sigma i}\right. 
      \left.+\frac{\zeta_1^2g_V^2}{3}\mathcal{O}_2\mathcal{G}(I)Y_{Vi}\right]$\\
      \hline
    &\multirow{1}{*}{$V_{B\bar{B}^{*}_2\to B\bar{B}^{*}_2}$}
    &$g_\sigma g_\sigma''\mathcal{O}_7Y_\sigma+\frac{1}{2}\beta\beta''g_V^2\mathcal{O}_7\mathcal{G}(I)Y_V$\\
    \hline
			
   &\multirow{3}{*}{$V_{B^{*}\bar{B}_1\to B^{*}\bar{B}_1}$}
   &$g_\sigma g_\sigma''\mathcal{O}_4Y_\sigma
   +\frac{5gk}{18f_\pi^2}\left(\mathcal{O}_5\mathcal{Z}+\mathcal{O}_6\mathcal{T}\right)\mathcal{H}(I)Y_P$&\\
		&&$+\frac12\beta\beta''g_V^2\mathcal{O}_4\mathcal{G}(I)Y_V
        -\frac{5}{9}\lambda\lambda''g_V^2\left(2\mathcal{O}_5\mathcal{Z}-\mathcal{O}_6\mathcal{T}\right)\mathcal{G}(I)Y_V$&\\
     &&$+c\times\left[\frac{h_\sigma^{\prime2}}{18\pi_\pi^2}\left[\mathcal{O}_5\mathcal{Z}+\mathcal{O}_6\mathcal{T}\right)Y_{\sigma i}\right. 
     \left.+\frac{\zeta_1^2g_V^2}{12}\mathcal{O}_5\mathcal{G}(I)Y_{Vi}\right]$\\
     \hline
      &\multirow{6}{*}{$V_{B^{*}\bar{B}^{*}_{2}\to B^{*}\bar{B}^{*}_{2}}$}
      &$g_{\sigma}g_{\sigma}''\frac{\mathcal{O}_{14}+\mathcal{O}_{14}'}{2}Y_{\sigma}$\\
      &&$+\frac{gk}{3f_{\pi}^{2}}\left(\frac{\mathcal{O}_{15}+\mathcal{O}_{15}'}{2}\mathcal{Z}+\frac{\mathcal{O}_{16}+\mathcal{O}_{16}'}{2}\mathcal{T}\right)\mathcal{H}(I)Y_{P}$&\\
    &&$+\frac{1}{2}\beta\beta''g_{V}^{2}\frac{\mathcal{O}_{14}+\mathcal{O}_{14}'}{2}\mathcal{G}(I)Y_{V}$\\
    &&$-\frac{2}{3}\lambda\lambda''g_{V}^{2}\left(2\frac{\mathcal{O}_{15}+\mathcal{O}_{15}'}{2}\mathcal{Z}-\frac{\mathcal{O}_{16}+\mathcal{O}_{16}'}{2}\mathcal{T}\right)\mathcal{G}(I)Y_{V}$&\\
    &&$+c\times\left[\frac{h_\sigma^{\prime2}}{3f_\pi^2}\left(\frac{\mathcal{O}_{17}+\mathcal{O}_{17}^{\prime}}2\mathcal{Z}+\frac{\mathcal{O}_{18}+\mathcal{O}_{18}^{\prime}}2\mathcal{T}\right)Y_{\sigma i}\right.$\\
    &&$\left.+\frac{\zeta_1^2g_V^2}2\frac{\mathcal{O}_{17}+\mathcal{O}_{17}^{\prime}}2\mathcal{G}(I)Y_{Vi}\right]$&\\
    \hline
	&\multirow{1}{*}{$V_{B_s\bar{B}_{s1} \to B_s\bar {B}_{s1}} $}
   &$\frac{1}{2}\beta\beta''g_V^2\mathcal{O}_1Y_\phi +c\times \left( \frac{\zeta_1^2g_V^2}{3}\frac{\mathcal{O}_2+\mathcal{O}_2^{\prime}}{2}Y_{\phi i}\right)$\\
\hline
     &\multirow{1}{*}{$V_{B_s\bar{B}_{s2}^{*} \to B_s\bar{B}_{s2}^{*}}$}
     &$\frac12\beta\beta''g_V^2\frac{\mathcal{O}_7+\mathcal{O}_7^{\prime}}2Y_\phi   $&\\
     \hline
      &\multirow{3}{*}{$V_{B_s^{*}\bar{B}_{s1} \to B_s^{*}\bar{B}_{s1}}$}
      &$\frac{5gk}{27f_\pi^2}\left(\mathcal{O}_5\mathcal{Z}+\mathcal{O}_6\mathcal{T}\right)Y_\eta $\\
&&$+\left(\frac{\beta\beta''g_V^2}2\mathcal{O}_4+\frac{5}9 \lambda\lambda''g_V^2 (\mathcal{O}_6\mathcal{T}-2\mathcal{O}_5\mathcal{Z})\right)Y_\phi $&\\
&&$+c\times \frac{\zeta_1^2g_V^2}{12}{\mathcal{O}_{5}}Y_{\phi i}$&\\
       \hline
      &\multirow{4}{*}{$V_{B_s^{*}\bar{B}^{*}_{s2} \to B_s^{*}\bar{B}^*_{s2}}$}
&$\frac{2}{9}\frac{gk}{f_\pi^2}\left(\frac{\mathcal{O}_{15}+\mathcal{O}_{15}^{\prime}}2\mathcal{Z}+\frac{\mathcal{O}_{16}+\mathcal{O}_{16}^{\prime}}2\mathcal{T}\right)Y_\eta$&\\
&&$+\left[\frac{2}{3}\lambda\lambda''g_V^2\left(\frac{\mathcal{O}_{16}+\mathcal{O}_{16}^{\prime}}{2}\mathcal{T}-2\frac{\mathcal{O}_{15}+\mathcal{O}_{15}^{\prime}}{2}\mathcal{Z}\right)\right.$\\
&&$\left.+\frac{1}{2}\beta\beta''g_V^2\frac{\mathcal{O}_{14}+\mathcal{O}_{14}^{\prime}}{2}\right]Y_{\phi}+c\times\left(\frac{1}{2}\zeta_1^2g_V^2\frac{\mathcal{O}_{17}+\mathcal{O}_{17}^{\prime}}{2}\right)Y_{\phi i}$&\\
		\end{tabular}
	\end{ruledtabular}
\end{table}

The relevant parameters in the present calculations are listed in Table~\ref{parameters}. To study the $H_{(s)}T_{(s)}$ and $H_{(s)}\bar{T}_{(s)}$ systems conveniently, one can define two functions, $\mathcal{H}(I)Y(\Lambda,m_P,r)$ and $\mathcal{G}(I)Y(\Lambda,m_V,r)$, in analogy to the doubly charm systems~\cite{Wang:2021yld}, that is,
\begin{eqnarray}
\mathcal{H}(0)Y(\Lambda,m_P,r)=-\frac{3}{2}Y(\Lambda,m_{\pi},r)+\frac{1}{6}Y(\Lambda,m_{\eta},r),
\end{eqnarray}
\begin{eqnarray}
\mathcal{H}(1)Y(\Lambda,m_P,r)=\frac{1}{2}Y(\Lambda,m_{\pi},r)+\frac{1}{6}Y(\Lambda,m_{\eta},r),
\end{eqnarray}
\begin{eqnarray}
\mathcal{G}(0)Y(\Lambda,m_V,r)=-\frac{3}{2}Y(\Lambda,m_{\rho},r)+\frac{1}{2}Y(\Lambda,m_{\omega},r),
\end{eqnarray}
\begin{eqnarray}
\mathcal{G}(1)Y(\Lambda,m_V,r)=\frac{1}{2}Y(\Lambda,m_{\rho},r)+\frac{1}{2}Y(\Lambda,m_{\omega},r).\label{vectorisopin}
\end{eqnarray}
Here, the $\mathcal{H}(I)$ and $\mathcal{G}(I)$ are the spin factors of the $H_{(s)}\bar T_{(s)}$ systems, where the $I$ is the isospin quantum number. For the $H_{(s)}T_{(s)}$ system, we have $\mathcal{H}'(I) = -\mathcal{H}(I)$ and $\mathcal{G}'(I) = -\mathcal{G}(I)$. Also, We define
$$\left.Y_i=
\begin{cases}
|q_i|\leqslant m,\frac{e^{-m_ir}-e^{-\Lambda_i^2r}}{4\pi r}-\frac{\Lambda_i^2-m_i^2}{8\pi\Lambda_i}e^{-\Lambda_ir}; \\
|q_i|>m,\frac{\cos(m_i^{\prime}r)-e^{-\Lambda_ir}}{4\pi r}-\frac{\Lambda_i^2+m_i^{\prime2}}{8\pi\Lambda_i}e^{-\Lambda_ir}; & 
\end{cases}\right.$$
where $m_i^2 =m^2 - q_i^2$, $m_i'^2 =q_i^2 - m^2$, $q_i^2=\left(\frac{M_A^2+M_D^2-M_C^2-M_B^2}{2(M_C+M_D)}\right)^2$, and $\Lambda_i^2 =\Lambda^2 - q_i^2$. The $M_A$, $M_B$, $M_C$, and $M_D$ represent the masses of the particles. The masses of elementary particles are compiled in the Particle Data Group review~\cite{ParticleDataGroup:2024cfk}. Additionally, the operators are defined as $\mathcal{Z}=\frac{1}{r^2}\frac{\partial}{\partial r}r^2\frac{\partial}{\partial r}$, $\mathcal{T}=r\frac{\partial}{\partial r}\frac{1}{r}\frac{\partial}{\partial r}$.
The operators $\mathcal{O}_k^{(\prime)}$ are defined following the conventions established in Refs.~\cite{Wang:2020dya, Wang:2021aql}.

Moreover, it should be mentioned that the $H_{({s})}\bar{T}_{({s})}$ and $H_(s) T_{(s)}$ systems exhibit only minor differences when higher-order terms are neglected. Hence, we only present the potential functions for the $H_{s}\bar{T}_{s}$ system in Table ~\ref{zb}. The explicit potentials for the $H_(s) T_{(s)}$ system can be found in Refs.~\cite{Wang:2021yld, Wang:2020dya}.


\end{document}